\documentclass[final, 5p,12pt]{elsarticle}
\pdfoutput=1
\usepackage{graphicx}
\usepackage{epsfig} 
\usepackage{amsmath}
\usepackage{times}
\usepackage{subcaption}

\begin{document}
\title{Propagation of Disturbances in AC Electricity Grids}

\author[add1,add2]{Samyak Tamrakar}
\author[add1]{Michael Conrath}
\author[add1,add3]{Stefan Kettemann}
\address[add1]{Jacobs University, Department of Physics and Earth Sciences, Campus Ring 1, 28759 Bremen, Germany}
\address[add2]{Institute of Physics, Carl von Ossietzky Universit\"at Oldenburg, Ammerl\"ander Heerstra\ss e 114-118, 26129 Oldenburg}


\address[add3]{Division of Advanced Materials Science, Pohang University of Science and Technology (POSTECH), San 31, Hyoja-dong, Nam-gu, Pohang 790-784, South Korea}
        
\begin{abstract} 
The energy transition towards high shares of renewable energy resources will affect the dynamics and the stability of electricity grids in many ways.  It is therefore crucial to understand its impact. We aim to contribute to this understanding by solving the  dynamic swing equations describing the coupled rotating masses of synchronous generators and motors. On  different grid topologies we identify parameter regimes with very different transient dynamics: the disturbance may either decay exponentially in time, superimposed by oscillations, with the fast decay rate of a single node, or  with a smaller decay rate without oscillations. Most remarkably, as the inertia is lowered, the nodes may become more correlated, slowing down the spreading of a disturbance, decaying  slowly with a power law in time. We show that this collective effect exists in meshed transmission grids, but is absent in  tree grids. We conclude by discussing consequences  for the stability of transmission grids
 if no measures are undertaken to substitute the inertia of conventional power plants.
\end{abstract}

\date{\today }
\maketitle


 In order to cover
   the increasing human energy demand by  renewable energy 
 resources and to 
     ensure that this energy will be  available 
       wherever and whenever it is  needed, 
      more 
        efficient energy transport and storage technologies need to be developed.
 The fluctuations in generated power by wind turbines and solar cells - both in time and geographically - demand to explore new strategies to store energy on all time scales and to distribute 
the power in the grid smartly.  At the same time, the spreading of critical disturbances throughout the grid has to be prevented to ensure the 
stability of the entire grid. 
    Renewable energy resources fluctuate strongly in time on 
     time scales as small as seconds. Moreover the 
     inverter-connected wind turbines and solar cells  
  provide no inertia \cite{ulbig}.  This is in contrast to  conventional generators, whose 
       rotating masses hold inertia and thereby  
        momentary power reserve available for the grid, which makes the 
         grid resilient and  prevents  strong 
         fluctuations of the grid  frequency on time scales 
        of  several seconds \cite{Kundur1994,Machowski2008}. 
           As the inertia in the grid keeps decreasing with higher 
            share of renewables, the 
         grid is responding on shorter time scales to disturbances. 
          It is therefore essential to understand the impact of this development on the stability of electricity grids. 
   In this article,  
we aim to find out if and 
 how  the propagation of disturbances in AC grids
is modified when the grid  inertia   from 
  the rotating masses of 
   generators is decreasing. 

The dynamic interaction and  response of generators and consumers
  is studied modeling the grid as a network of nonlinear oscillators \cite{Kundur1994,Machowski2008,Bergen1981,Filatrella2008,Rohden2012,Schmietendorf2014}.    These 
  nonlinear swing equations describe the dynamics of  coupled 
   rotating masses by  a system of coupled differential equations for local rotor angles  $ {\varphi_i , } $  where  $ {i } $ denote the grid nodes. 
As we aim to contribute to the  understanding
 of   how disturbances evolve with time in AC grids,
 we consider   control free grids without primary and secondary control measures \cite{ulbig,Kundur1994,Machowski2008}. The origin of  disturbances can be fluctuations in generating power or  sudden changes of transmission line capacitance. We solve the nonlinear dynamic power balance 
equations numerically  and  explore how a local 
 perturbation propagates with time throughout  the grid. 
 We analyze these  numerical results, employing 
 analytical results as obtained from  mapping  the swing equations on  discrete linear wave equations  for  small perturbations \cite{Kettemann2016}. 
  Depending on the geographical distribution of power, grid power transmission capacity and topology we find that the disturbance  may either  decay exponentially in time
   with the decay rate of a single oscillator  $ {\Gamma^0 , } $  or exponentially 
   with a smaller decay rate  $ {\Gamma < \Gamma^0 , } $  
   or,  even more slowly, decaying  with a power law in time. 
     Such a slow power law  decay 
  arise together with a slow,  diffusive  propagation \cite{Kettemann2016}. 
       

\section*{Results}
\begin{figure*}[ht]
\centering
	\begin{subfigure}{0.3\textwidth}
		\includegraphics[width = \textwidth]{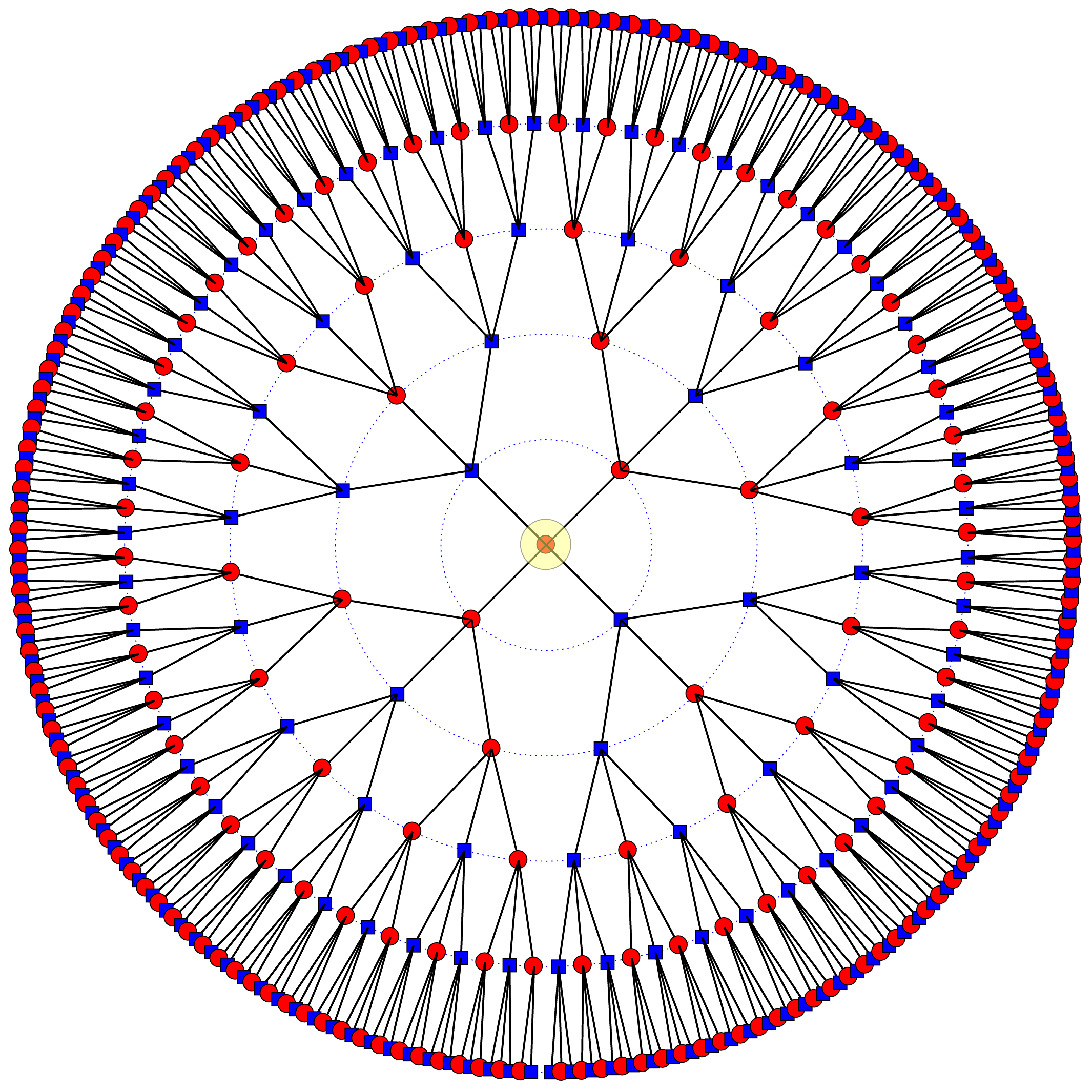}
		\caption{}
	\end{subfigure}
	\begin{subfigure}{0.3\textwidth}
		\includegraphics[width = \textwidth]{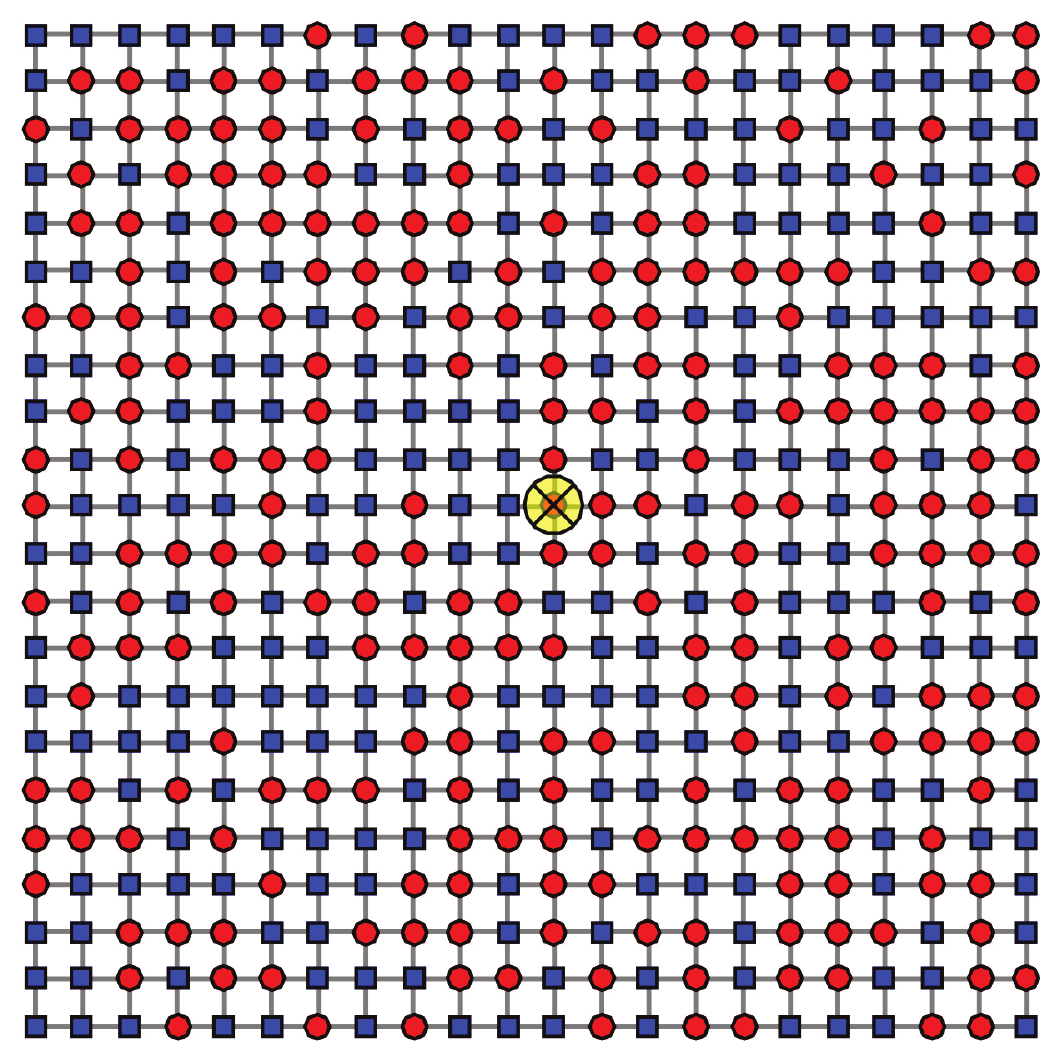}
		\caption{}
	\end{subfigure}
	\begin{subfigure}{0.3\textwidth}
		\includegraphics[width = \textwidth]{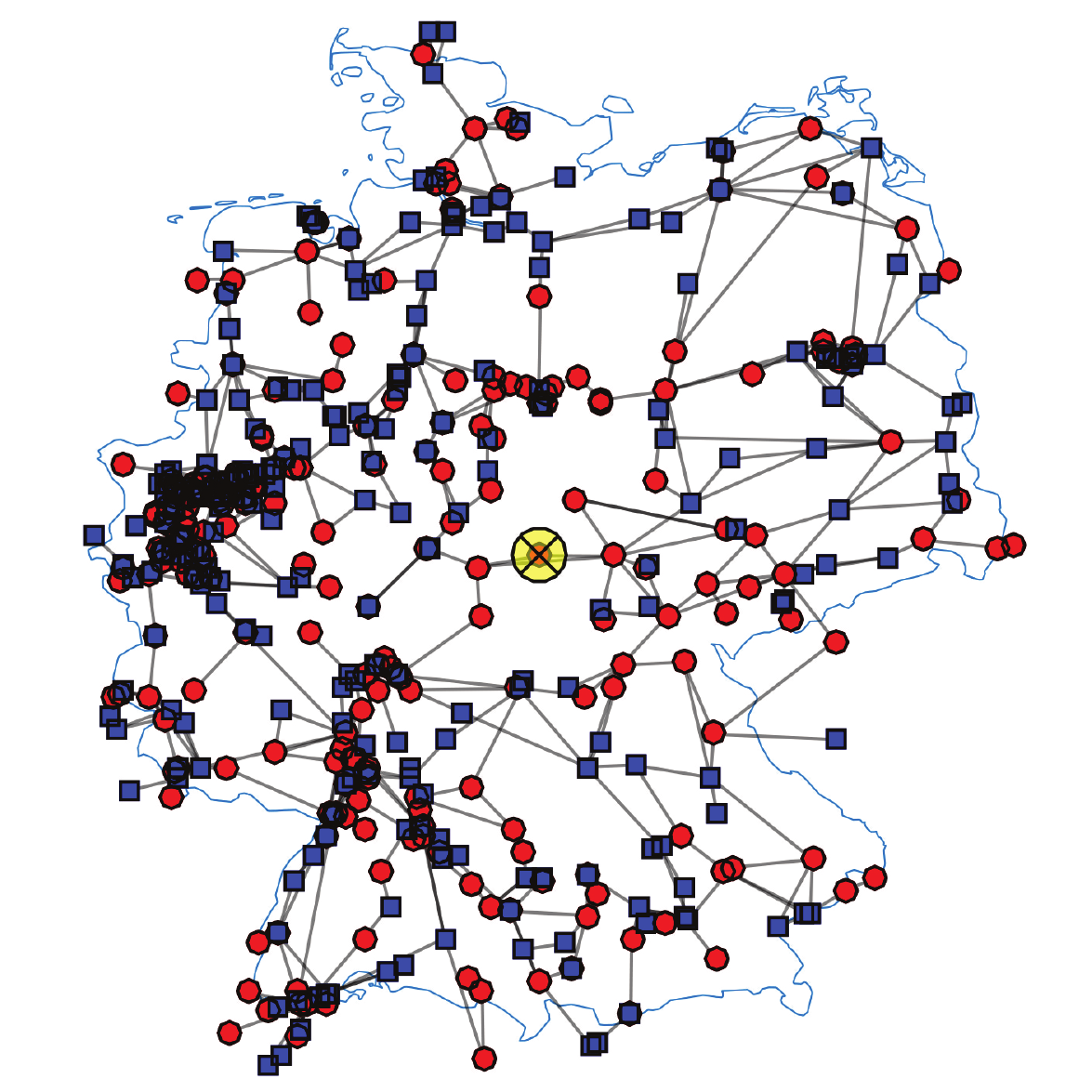}
		\caption{}
	\end{subfigure}
	\caption{Grid topologies under consideration in the present study. (a) Cayley tree grid with 5 branching generations and 3 new branches `growing' from every point giving  $ {N=484 } $ nodes and 483 links. (b) 22x22 square grid with  $ {N= 484 } $ nodes, 924 links and random arrangement of generators and consumers. (c) German transmission grid model with  $ {N=502 } $ nodes and 673 links \cite{scigridv2}. The parameters of square and tree grid, respectively, have been chosen in favor of comparable size to the German grid. Red circles represent generators, blue squares represent motors and the yellow circle with `X' marks the spot where a perturbation is applied.}\label{fig:grid}
\end{figure*}

\begin{figure}
\centering
 \includegraphics[width = 0.8\columnwidth]{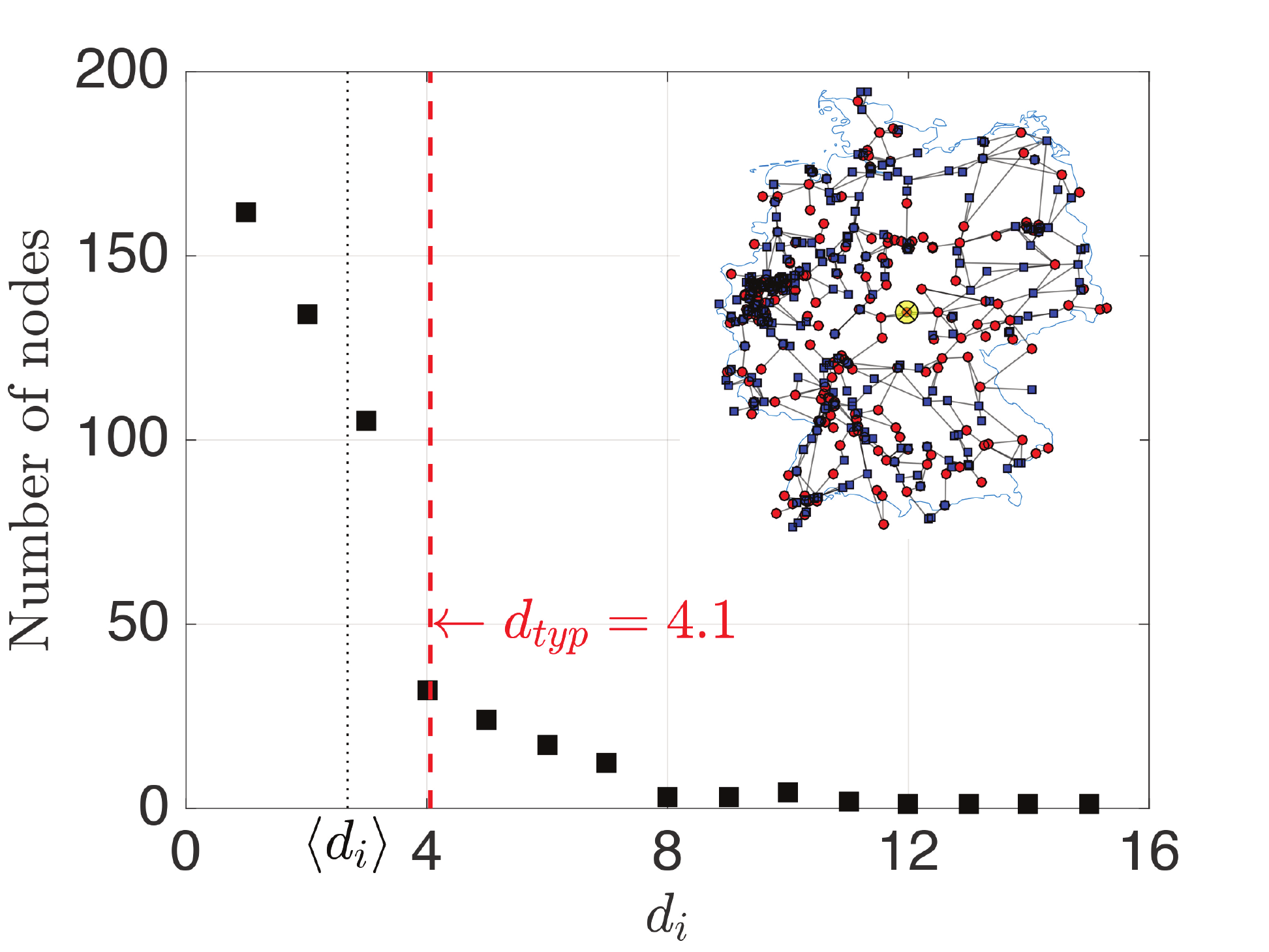}
 \caption{Distribution of  node degree  $ {d_i } $ for German transmission grid.} 
 \label{fig:degree}
\end{figure}

{\bf Grid Topologies.} Aiming at a systematic approach, we consider three different grid topologies.
Firstly, the Cayley tree 
grid,  Fig.  \ref{fig:grid}  a),  resembles typical distribution grids which are preferably operated in such a tree-like fashion to pinpoint and repair failures more easily. Starting from a center, a constant number of branches grows 
outward in a given number of branching levels, each time with the same branching number
  $ {b . } $  Hence, the tree grids are 
 characterized by   $ {b } $ and 
 the level  $ {l } $ of branching, the distance between neighbored nodes
 $ {a } $ and the total number of nodes  $ {N . } $ 
 An important characteristic of grids is the degree  $ {d_i , } $  the number of links 
  connecting node  $ {i } $ to any other node. 
 The degree of this
  tree  grid is  $ {d = b+1 , } $  with  the exception of the edge nodes, which have degree  $ {d=1 . } $ 
 Secondly, the square grid,  Fig. \ref{fig:grid} b). It is a simple meshed topology, used as basic model for transmission grids with their 
strict redundancy demand to guarantee continuing operation even when a single line fails (n-1 criterion). These square grids 
are characterized by the distance between neighbored nodes
 $ {a , } $  the   linear grid size   $ {L } $  and the number of nodes  $ {N=(L/a)^2 . } $ 
  Their degree is constant,   $ {d =4 , } $  except at  edges ( $ d=3 $ ) and corners ( $ d=2 $ ). 
Thirdly,  we choose the German 
transmission grid, Fig. \ref{fig:grid} c)  as a real-world example with an irregular realistic topology, which is inhomogeneous  and highly meshed. As reference for the German transmission grid, the open-source SciGRID dataset was used \cite{scigridv2}. Excluding island nodes in it, the largest 
connected network of the four highest voltage levels,  400 kV, 380 kV, 220 kV
as well as some 110 kV lines,   was adopted as grid model. It comprises  $ {N = 502 } $ nodes and 673 links.
It has  a wide  distribution  of  degree  $ {d_i , } $  Fig.  \ref{fig:degree},
 with average degree   $ {\langle d_i \rangle = 2.7} $  and typical degree   $ {d_{typ}
    = \sqrt{ {d_i}^2 } = 4.1 . } $  Excluding  stubs, 
   singly connected nodes,  mostly due to the artifact that  this data set does not include the transnational European grid,  we get an average degree 
    $ {\langle d_i \rangle = 3.5 . } $  

{\bf Dynamic AC Grid Model.}
 3-phasic alternating current (AC) electric grids can be modeled by active power balance equations \cite{Bergen1981,Filatrella2008,Rohden2012,Heuck2013,Rohden2014,Menck2014}. Since the three phases are typically loaded and operated symmetrically, they can 
effectively be reduced to a single phase. Furthermore, linking the grid nodes, purely inductive lines are assumed, neglecting the  Ohmic losses along the lines,
 which are relatively small in  high voltage transmission grids. The combined inductance of an individual 3-phasic line between nodes 
 $ {i } $ and  $ {j } $ is  $ {L_{ij} . } $  The line  power capacity is then 
 $ {K_{ij}=|V_i| |V_j|/(\omega L_{ij}) } $ where  $ {V_i , } $   $ {V_j } $ 
 are the voltages at node  $ {i } $ and  $ {j . } $ 
   $ {\omega } $ is related to the     grid frequency  $ {f= 50\mathrm{~Hz} } $ by   $ {\omega=2\pi f . } $   We  will focus on the dynamics of the phases  $ {\varphi_i, } $ 
while the voltage amplitudes 
 are taken constant, since they typically  change only on larger time scales. 
    We assume the same voltage amplitude 
       throughout the grid, so that   $ {V_i=V \exp(\i \phi_i) } $ with  $ {\phi_i } $ being the voltage phase. Since our main goal is to study the influence of the grid topology and the inertia on the phase dynamics
  we  assumed  that all links have equal inductance  $ {L } $ and thus 
   equal power capacity  $ {K = V^2/(\omega L) . } $ 
   Thus,  the power capacity linking nodes  $ {i } $ and  $ {j } $ is given by 
     $ {K_{ij}=K A_{ij} , } $  where  $ {A_{ij} } $ is the 
adjacency matrix of the grid.
   Under these conditions, the stationary active power flow in the grid is 
    obtained from Kirchhoff's laws as function of 
the voltage phases  $ {\phi_i } $ as
\begin{equation}\label{Eq:grid_balance}
    P_i= K \sum_j A_{ij} \sin(\phi_i-\phi_j)~~,~~\sum P_i=0,
\end{equation}
with the second equation being the power balance condition,
 which has to be preserved at any time in order to avoid frequency shifts and is therefore controlled by 
  grid operators.  However, the nodal phases 
   vary with time.  The constant grid 
    frequency gives a phase shift which increases linearly in time as    $ {\omega t , } $   
   independently of the node  $ {i . } $ 
    Denoting the solution of the   stationary balance equation Eq. (\ref{Eq:grid_balance})
     by the  phase 
shift  $ {\theta_i^0 , } $  we can write 
\begin{equation} \label{Eq:phi}
    \phi_i(t)=\omega t+\theta_i^0 +\alpha_i(t)~~,
\end{equation}
with  the additional  dynamic, time dependent phase shift denoted by  $ {\alpha_i(t) . } $  
In the present work all grid nodes  are assumed to be connected to rotating machines,  either synchronous  generators or motors with inertia  $ {J_i . } $  All rotating machines and therefore all nodes are assigned an equal  magnitude of 
 electric power  $ {P , } $    which is either positive (generator) or 
negative (motor). Consequently,  $ {P_i=s_i P , } $ 
 with  $ {s_i \in  \{+,-\} } $ and  $ {J_i=J . } $  The phase dynamics is then governed by the balance of the change in kinetic  energy, the energy dissipation in the machines, and the electrical energy exchange with adjacent grid nodes,
 yielding the swing equations \cite{Bergen1981,Filatrella2008,Rohden2012}.
\begin{equation}\label{Eq:Pi_coupling}
 P_i=\left(\frac{J}{2}\frac{\mathrm{d}}{\mathrm{d}t}+\gamma\right)\left(\frac{\mathrm{d}\phi_i}{\mathrm{d}t}\right)^2+\sum_j F_{ij},
\end{equation}
where  $ {\gamma } $ is the damping constant and \newline ${F_{ij} = K_{ij} \sin(\phi_i-\phi_j)} $ 
 is the power flow in the transmission line between nodes  $ {i } $ and  $ {j . } $  
We note that, if  the nodes would not be coupled by the transmission lines, the phase  at each  node would decay exponentially fast  with the local 
 relaxation time   $ {\tau = J/\gamma , } $  increasing with inertia  $ {J } $ and decreasing with 
damping parameter  $ {\gamma . } $  Therefore, when studying the effect of the coupling between the nodes, it is convenient to scale the time with relaxation time  $ {\tau . } $  
 In order to study the spreading of perturbations,
 the temporal and spatial dependence of   $ {\alpha_i , } $  
  we insert  Eq. (\ref{Eq:phi}) into Eq. (\ref{Eq:Pi_coupling})
and  for small phase   velocities,  $ { \partial_t \alpha_i \ll \omega, } $ 
we arrive at the swing equations in dimensionless form,
\begin{equation} \label{Eq:alpha_norm}
\begin{split}
 \tau^2  \partial_t^2  \alpha_i + 2 \tau  \partial_t  \alpha_i &= s_i \Pi_{P} - \Pi_K \sum_j  A_{ij} \\
  &\quad \times \sin (\theta_i^0-\theta_j^0 + \alpha_i-\alpha_j),
\end{split}
\end{equation} 
with dimensionless parameters
 $ {\Pi_P = J P/(\gamma^2 \omega) } $  and
  $ {	\Pi_K =J K/(\gamma^2 \omega) . } $

{\bf Transient Dynamics.}
Now, we proceed to   solve  the nonlinear swing equations Eq. (\ref{Eq:alpha_norm})
on the different grid topologies shown in Fig. \ref{fig:grid}
 for a   set of parameters   $ {(\tau, \Pi_K,  \Pi_{P}), } $  
in order to 
study the transient behavior 
of the AC grid when it  is perturbed by a local disturbance. 
\begin{figure}[ht!] \label{Fig:phase_perturbation}
\centering
  \includegraphics[width =0.8\columnwidth]{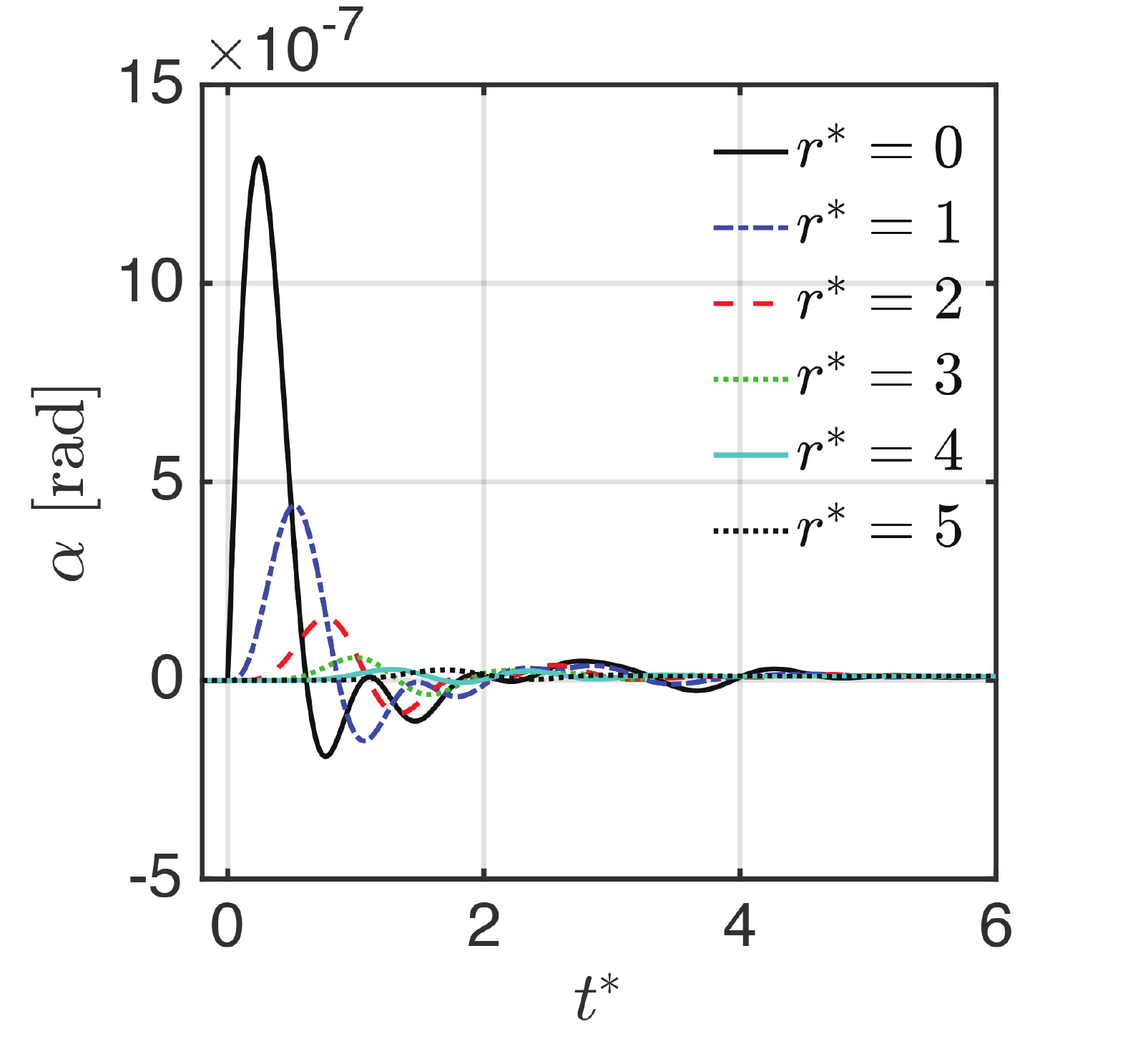}
  \caption{Phase perturbation  $ {\alpha(r^{*} = r/a, t^* = t/\tau) } $ for the 
   $ {b=3 } $ Caylee tree grid at nodes on different distances   $ {r=l a } $ away from the center node.}
  \label{fig:transient}
\end{figure}
As disturbance  of the stationary state we 
increase the power for a short time interval  
     $ {0\leq t \leq \Delta t_{pert} , } $  at the grid node marked with  `x' 
in the different grid topologies, 
Fig. \ref{fig:grid}.  We choose it as
one per mille of the initial power  $ {P , } $  corresponding to a perturbation energy
  $ {E_{pert}^{0}=0.001 \Pi_{P} \cdot \Delta t_{pert}. } $ 
 The resulting transient behavior of the phase deviation  $ {\alpha(t)} $ 
  is shown exemplary in Fig. \ref{fig:transient} for the  Cayley tree  grid of  Fig. \ref{fig:grid} a)
  as function of rescaled time  $ {t^* = t/\tau } $ for the 
  parameters  $ {(\Pi_K=10, \sigma  = \Pi_{P} / \Pi_{K} = 0.08) . } $  
  We see that  the phase at the disturbed  node,  $ {r^{*}=0 } $ (The distance
between any two nodes  $ {r_{ij} } $ is defined as the number of edges in a shortest path
connecting them \cite{Newman};  $ {r^{*} = r / a . } $ ) is 
     perturbed  first, reaching a maximum after a delay time, and then decaying in an oscillatory manner. As expected, phases of nodes further away from the 
      origin of the disturbance are perturbed later and reach smaller amplitudes. 
      We will analyze this temporal and spatial behavior quantitatively in the following. 
      \begin{figure}[h!] 
 \vspace{0cm}
 \centering
  \includegraphics[width =0.8\columnwidth]{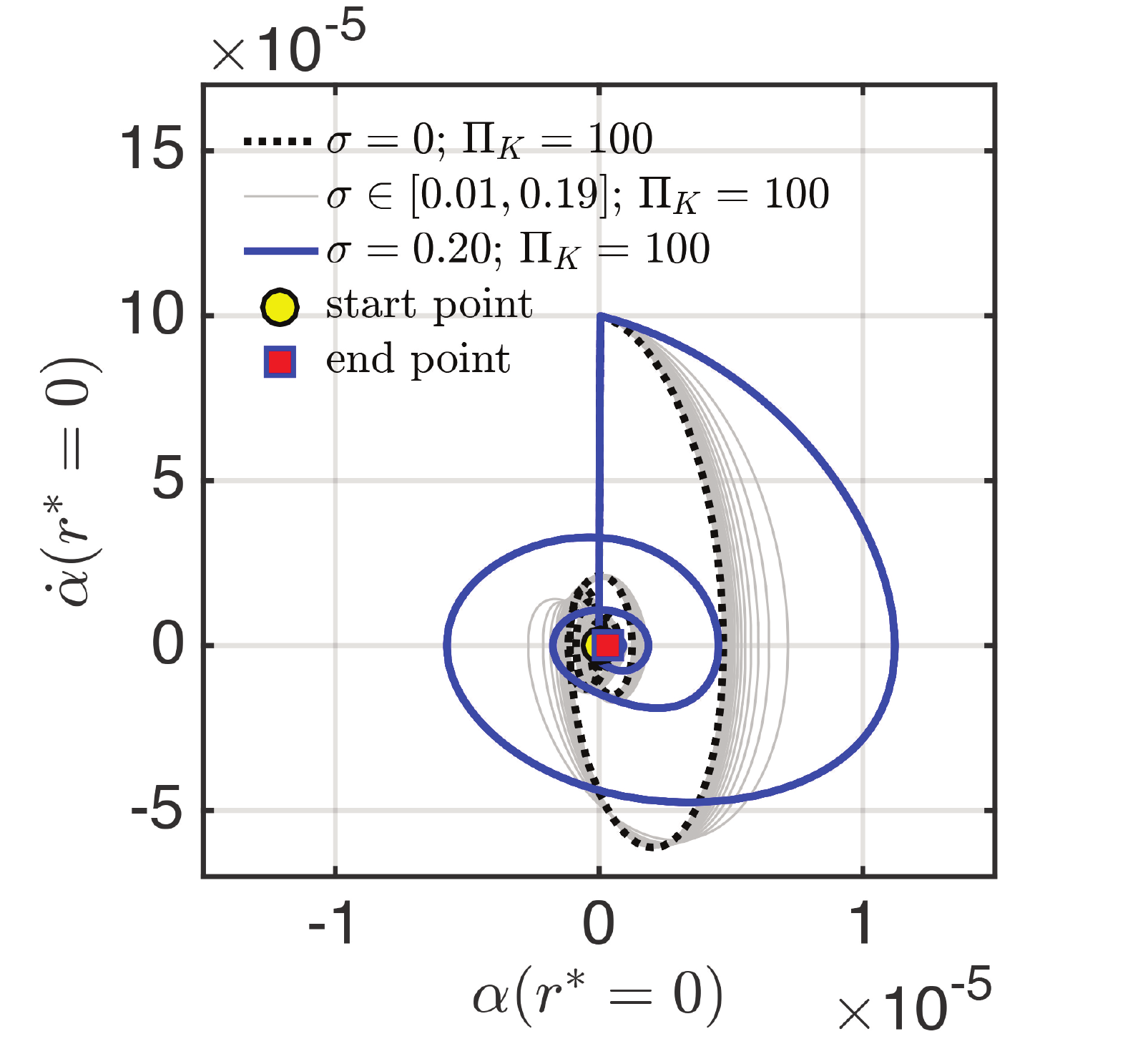}
  \vspace{0cm}
    \includegraphics[width =0.8\columnwidth]{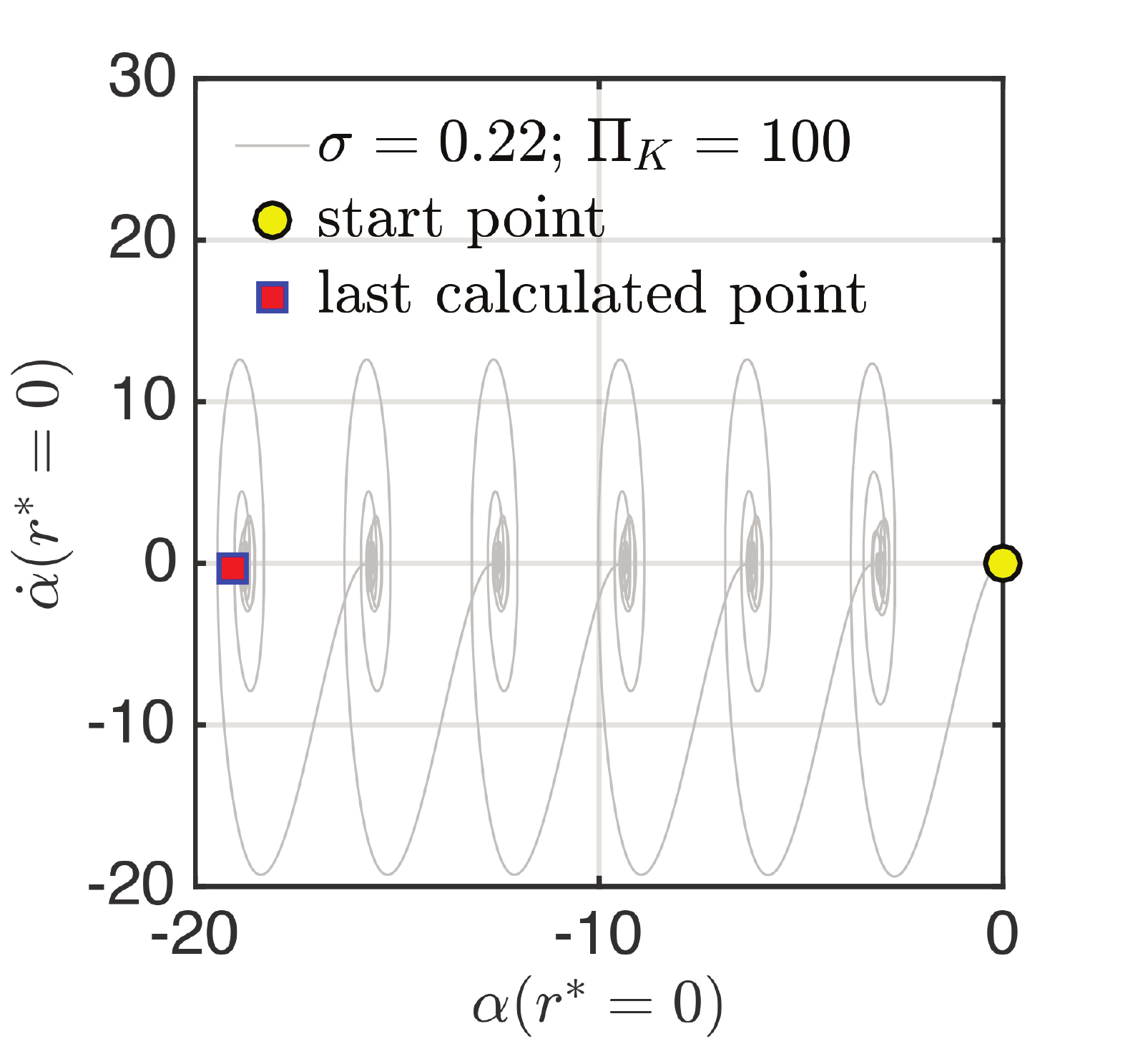}
    \vspace{0cm}
  \caption{Phase portrait    $ {\dot{\alpha}(r,t^*) } $ versus   $ {\alpha(r,t^*) } $  for the 
   $ {b=3 } $  Caylee tree grid for  $ {\Pi_{K} = 100 } $ and various  $ {\sigma . } $ }
  \label{fig:pp}
\end{figure}
         
Another way  to visualize this transient behavior  is the phase 
portrait  shown  exemplary in Fig. \ref{fig:pp},  a plot of  phase velocity  $ {\dot{\alpha_i} } $ versus 
phase  $ {\alpha_i . } $ 
 We see in the upper figure that the disturbance remains  within the basin of attraction of the attractive fixed point
   for   $ {\Pi_K = 100 } $ and  $ {\sigma } $ not exceeding  $ {0.2 . } $  There is a slight shift of the phase to which the perturbation decays at the end, which we find  
   to be due to  a global phase shift of all nodes induced by the disturbance at the center node. 
   In the lower figure it is seen that the perturbation destabilizes the grid when 
     $ { \sigma = 0.22, } $ where  the 
     phase deviation increases unbounded.
     Since we are interested in the propagation of small disturbances
     which do not destabilize the system, 
     we 
        review next  the conditions for stability in order 
       to   choose the size of the disturbance accordingly.

{\bf Stability.} 
           Not all values of the parameters
             $ {\Pi_P } $ and  $ {\Pi_K } $ allow for a stable stationary state. 
             Without any disturbance, the criterion to distinguish between stable (allowed) and unstable (forbidden) parameters is the existence of a non-complex solution for the 
stationary state, Eq. (\ref{Eq:grid_balance}).
Thus, the  ratio  $ {\sigma=\Pi_{P}/\Pi_K } $ determines whether parameters are allowed or forbidden and
a critical value  $ {\sigma_c } $  exists depending on  grid topology and power distribution.
If the power distribution is  $ {P_i = s_i P , } $   with  $ {s_i \in  \{+,-\} , } $ 
and 
there are no clusters of consumers or clusters of generators, the critical value at node  $ {i } $ is 
 given by   $ {\sigma_{c i}= d_i , } $  where  $ {d_i } $ is the node degree. Thus, the critical value 
  below which all nodes of the whole 
  grid are stable  is given by  $ {\sigma_{c}= {\rm min } (d_i) . } $ 
   For a general distribution of  $ {P_i , } $  there can be clusters of 
    generators or consumers. Then, 
    the critical value  $ {\sigma_{c}} $ 
      is determined by the size and form of the clusters.
      If  a cluster  of generators has the  total power 
     $ {P_{C} = \sum_{\rm cluster} P_i, } $ 
   with an effective degree  $ {d_C, } $  as obtained by 
    counting the number of consumers which are directly coupled to that cluster, 
     the 
   critical value of    $ {\sigma=\Pi_{P i}/\Pi_K} $  above which no stable solution exists, is given by 
      $ {\sigma_{c} = {\rm min }(P ~ d_C/P_{C}). } $ 
     
  Depending on the magnitude of the disturbance it 
   can destabilize the grid already  at   $ { \sigma < \sigma_c . } $   In the Suppl. 
 I. we  derive a typical upper limit for the size of the perturbation 
        $ {\alpha } $ before it kicks the system out of the stable region.
              We find that   the disturbance 
              can destabilize the grid at smaller values  $ { \sigma^* (\alpha) < \sigma_c . } $ 
               $ { \sigma^* (\alpha)} $ 
               coincides 
               with   $ {\sigma_c } $ only in the limit,
                when the perturbation amplitude 
                 is vanishing,  $ { \sigma^* (\alpha \rightarrow 0) = \sigma_c . } $ 

\begin{figure*}
	\centering
	\begin{subfigure}[b]{0.32\textwidth}
		\includegraphics[height = 0.775\textwidth]{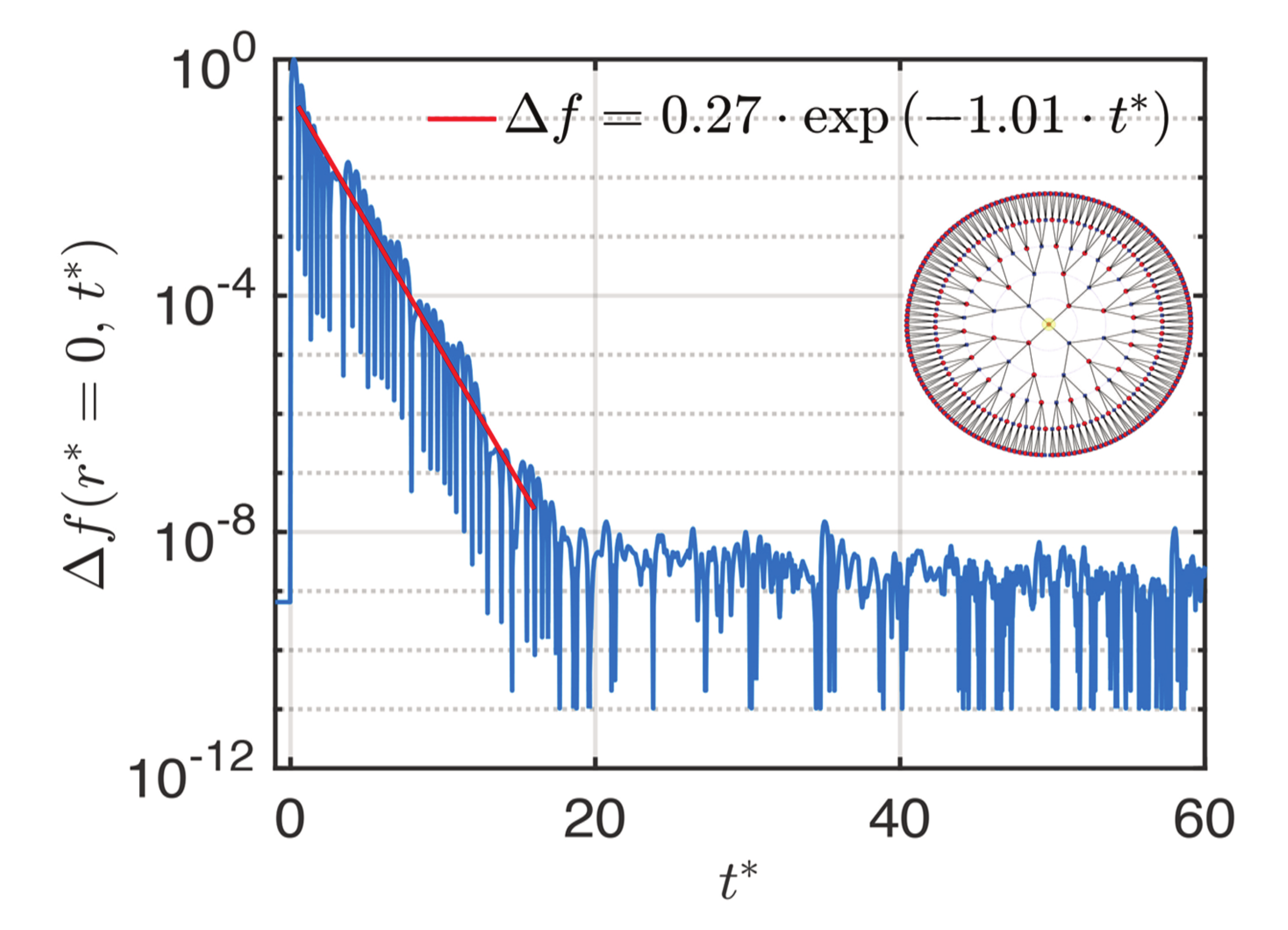}
		\caption{ $ {\sigma = 0.1 } $  and  $ {\Pi_{K} = 10} $ }
	\end{subfigure}
	\hspace{.15cm}
	\begin{subfigure}[b]{0.32\textwidth}
		\includegraphics[height = 0.8\textwidth]{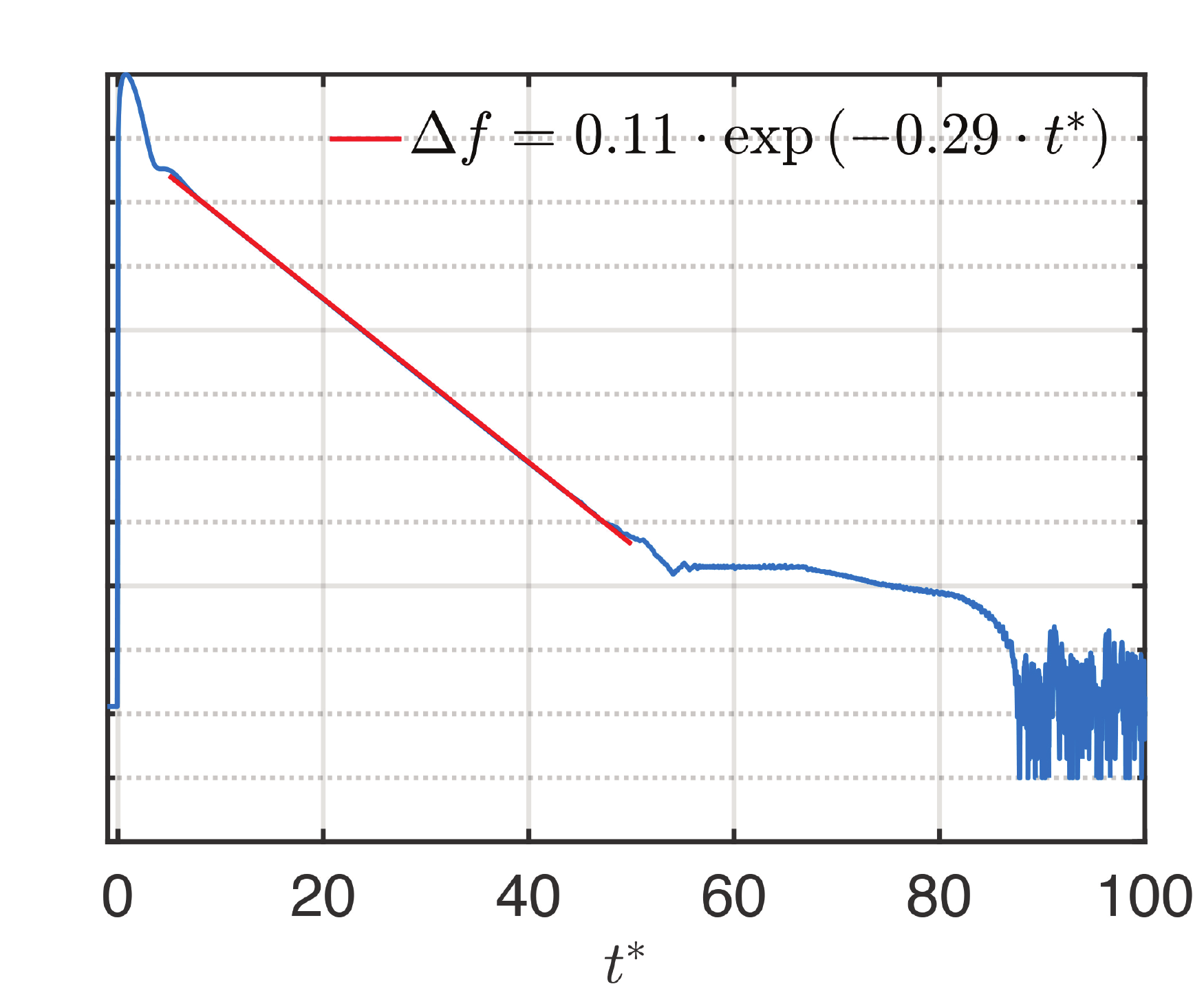}
		\caption{ $ {\sigma = 0.1 } $  and  $ {\Pi_{K} = 0.5} $ }
	\end{subfigure}
	\begin{subfigure}[b]{0.32\textwidth}
		\includegraphics[height = 0.8\textwidth]{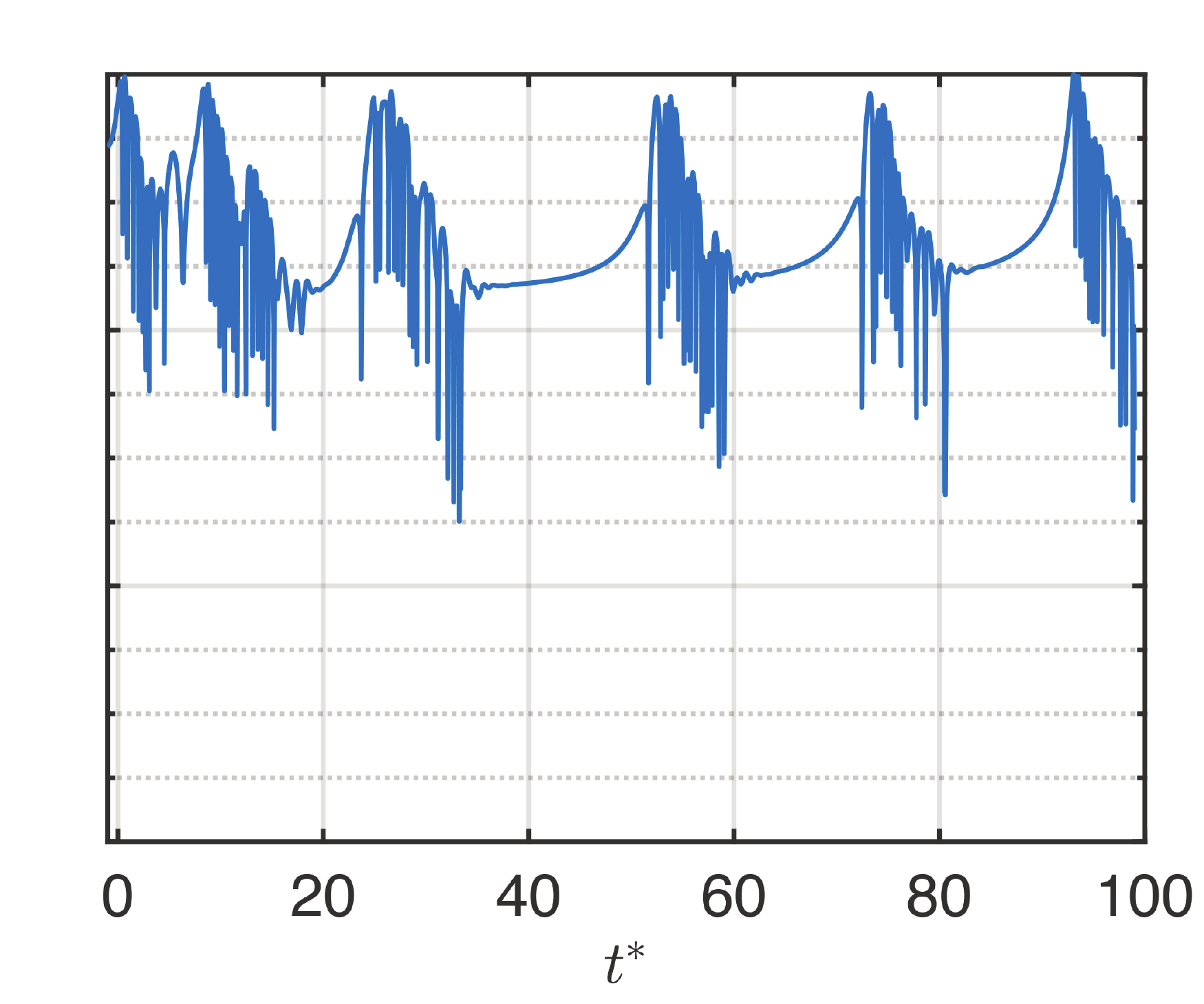}
		\caption{ $ {\sigma = 0.3 } $  and  $ {\Pi_{K} = 10} $ }
	\end{subfigure}
	\begin{subfigure}[b]{0.32\textwidth}
		\includegraphics[height = 0.78\textwidth]{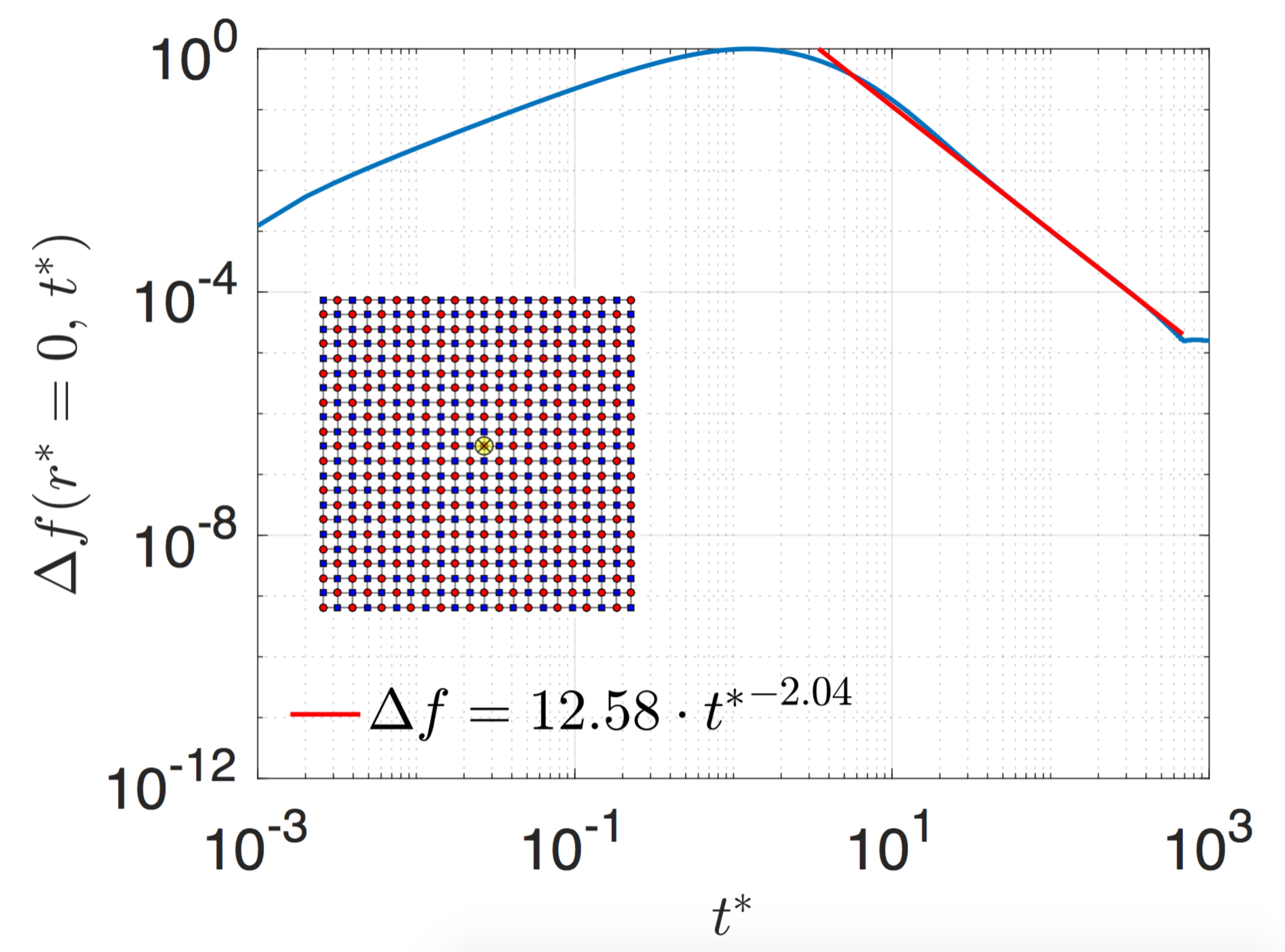}
		\caption{ $ {\sigma = 0.1 } $  and  $ {\Pi_{K} = 10^{-1}} $ }
	\end{subfigure}
	\hspace{.15cm}
	\begin{subfigure}[b]{0.32\textwidth}
		\includegraphics[height = 0.8\textwidth]{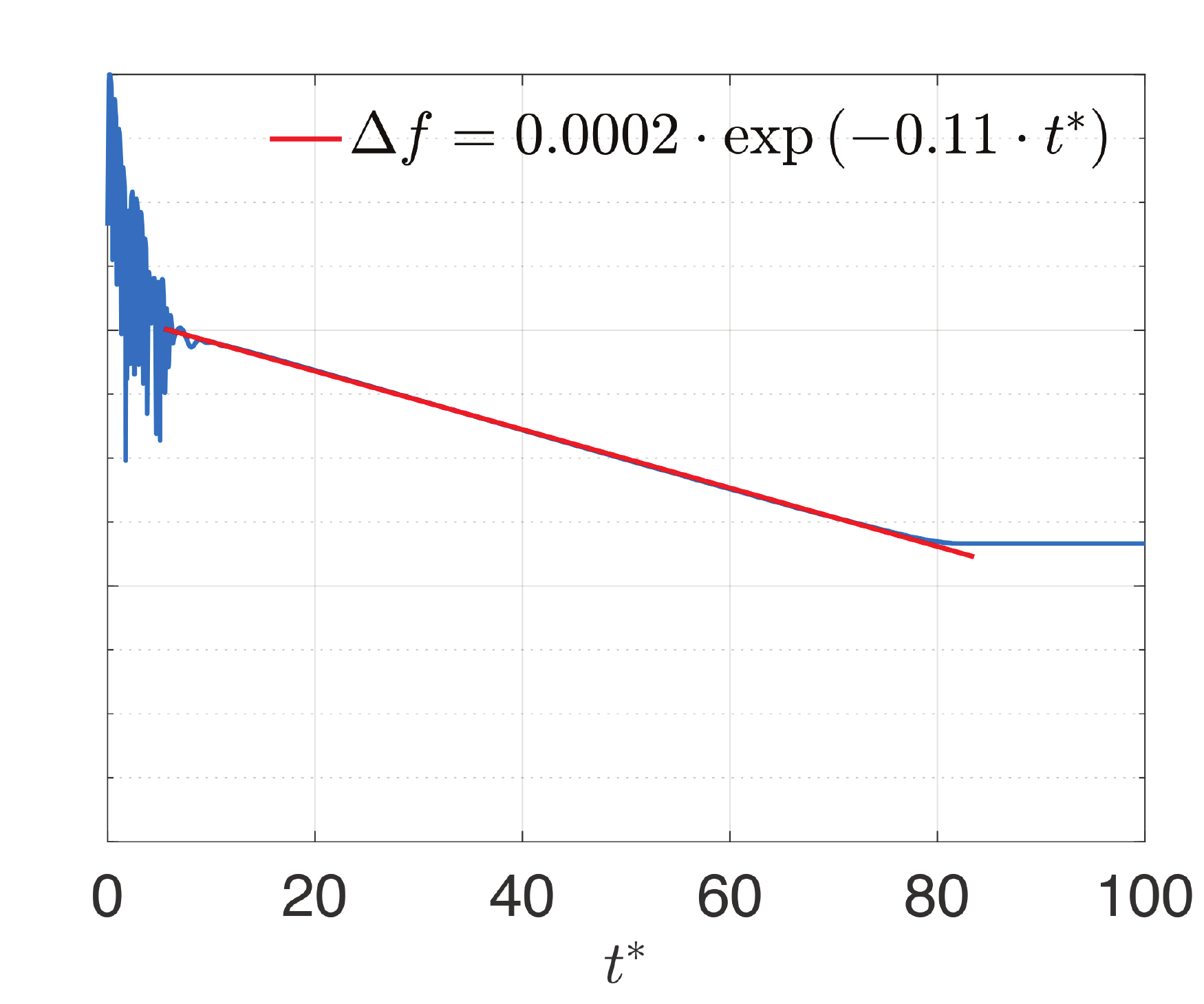}
		\caption{ $ {\sigma = 0.1 } $  and  $ {\Pi_{K} = 10} $ }
	\end{subfigure}
	\begin{subfigure}[b]{0.32\textwidth}
		\includegraphics[height = 0.8\textwidth]{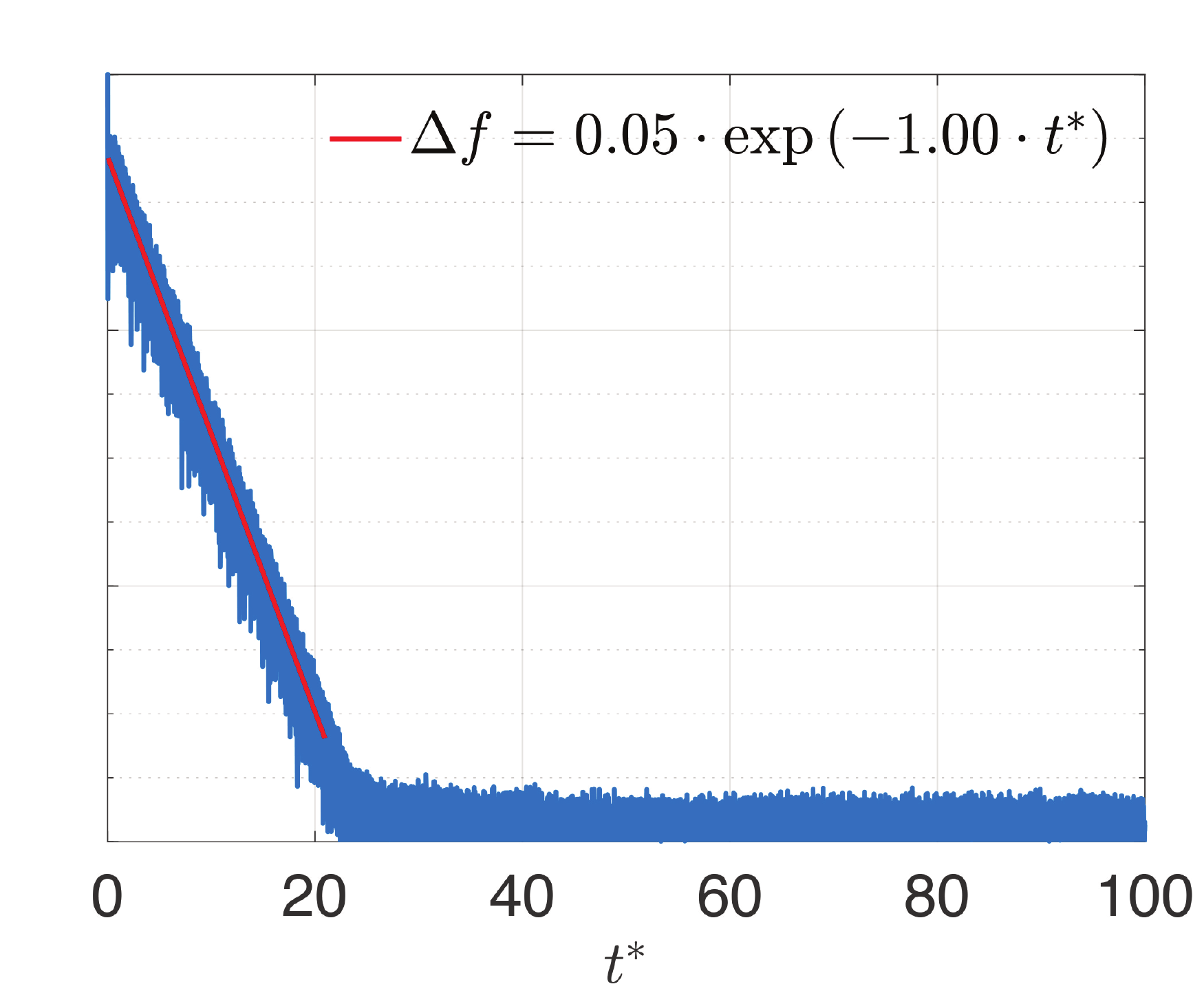}
		\caption{ $ {\sigma = 0.1 } $  and  $ {\Pi_{K} = 10^{5}} $ }
	\end{subfigure}
	\begin{subfigure}[b]{0.32\textwidth}
		\includegraphics[height = 0.79\textwidth]{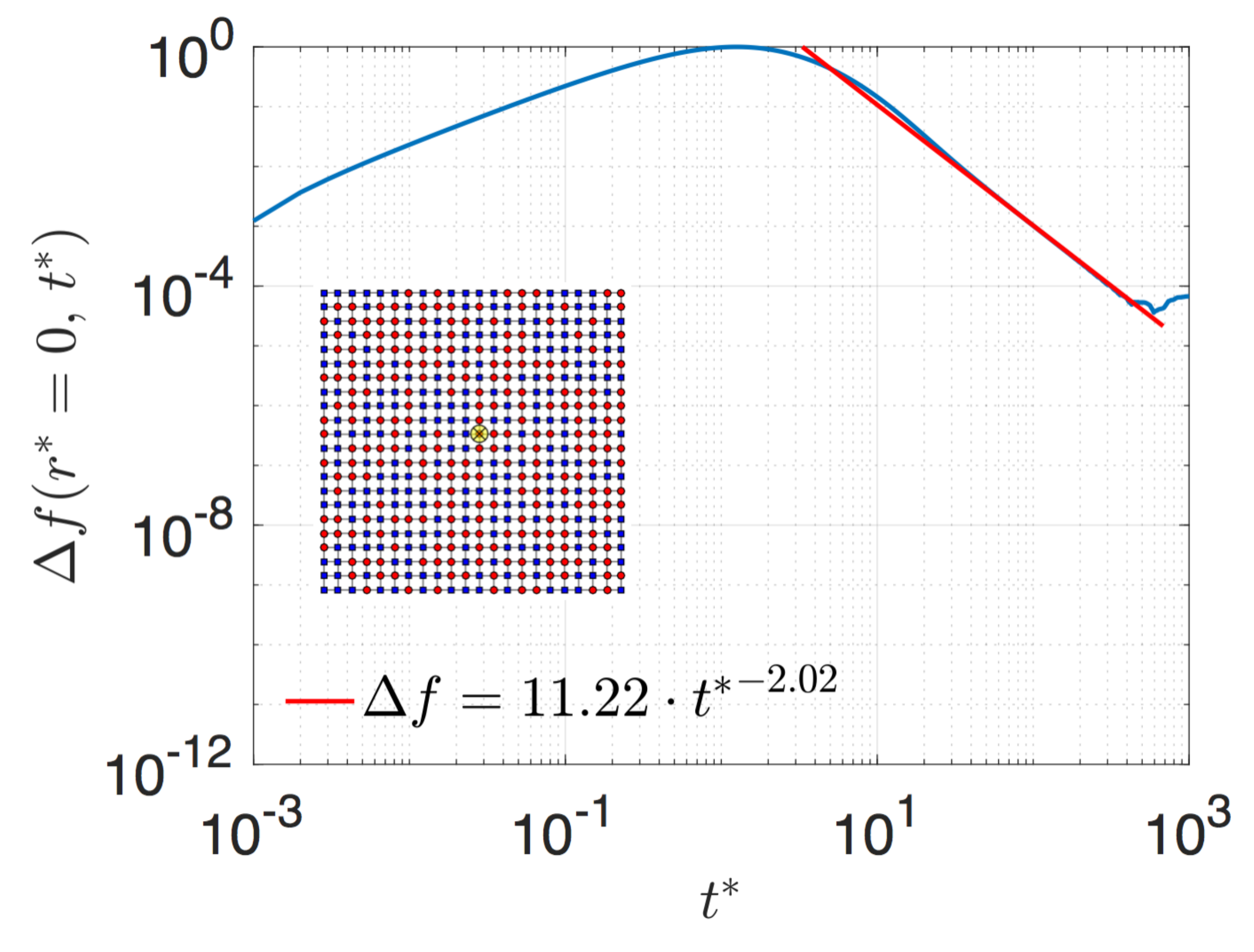}
		\caption{ $ {\sigma = 0.1 } $ and  $ {\Pi_{K} = 10^{-1}} $ }
	\end{subfigure}
	\hspace{.15cm}
	\begin{subfigure}[b]{0.32\textwidth}
		\includegraphics[height = 0.8\textwidth]{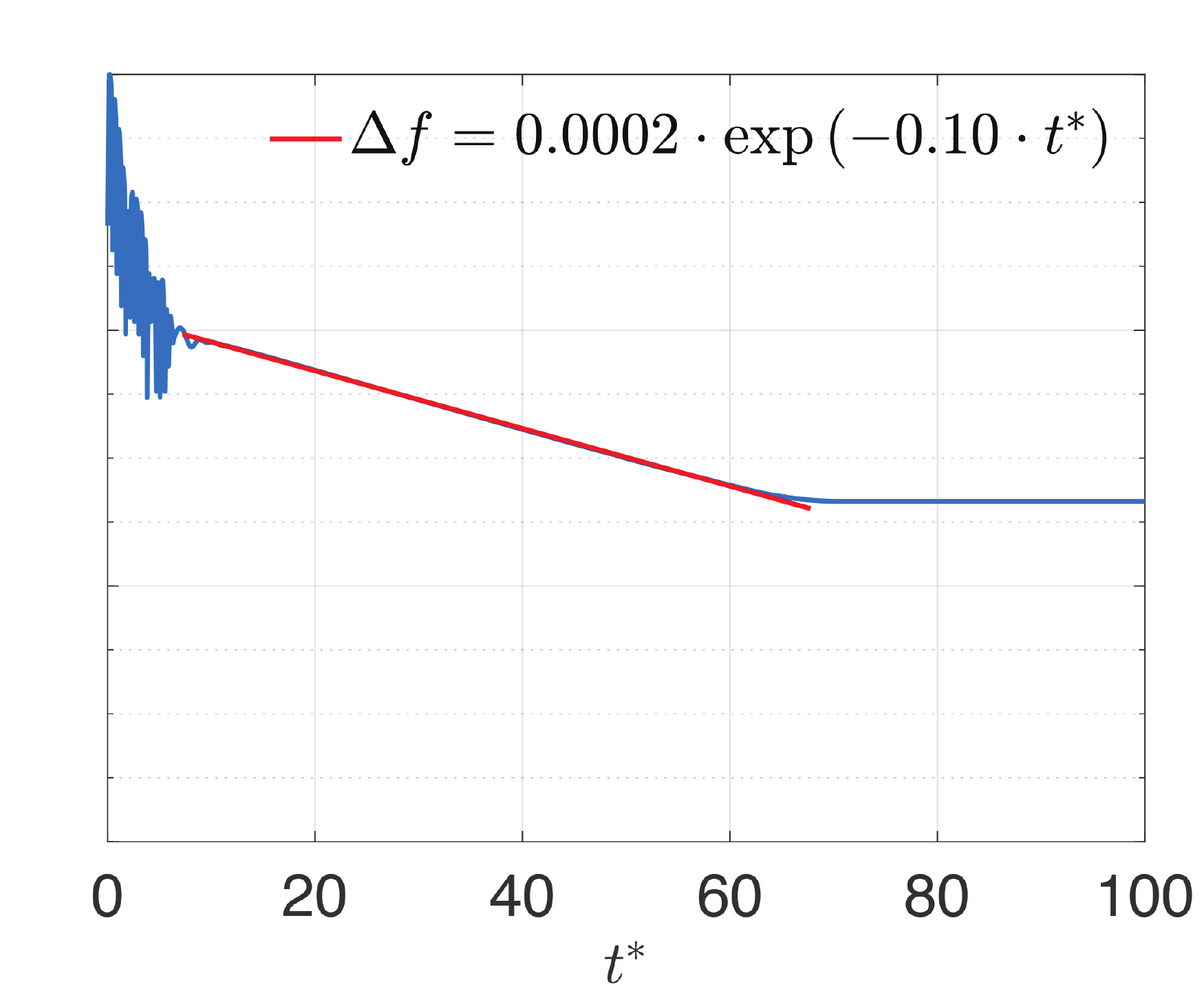}
		\caption{ $ {\sigma = 0.1 } $ and  $ {\Pi_{K} = 10} $ }
	\end{subfigure}
	\begin{subfigure}[b]{0.32\textwidth}
		\includegraphics[height = 0.8\textwidth]{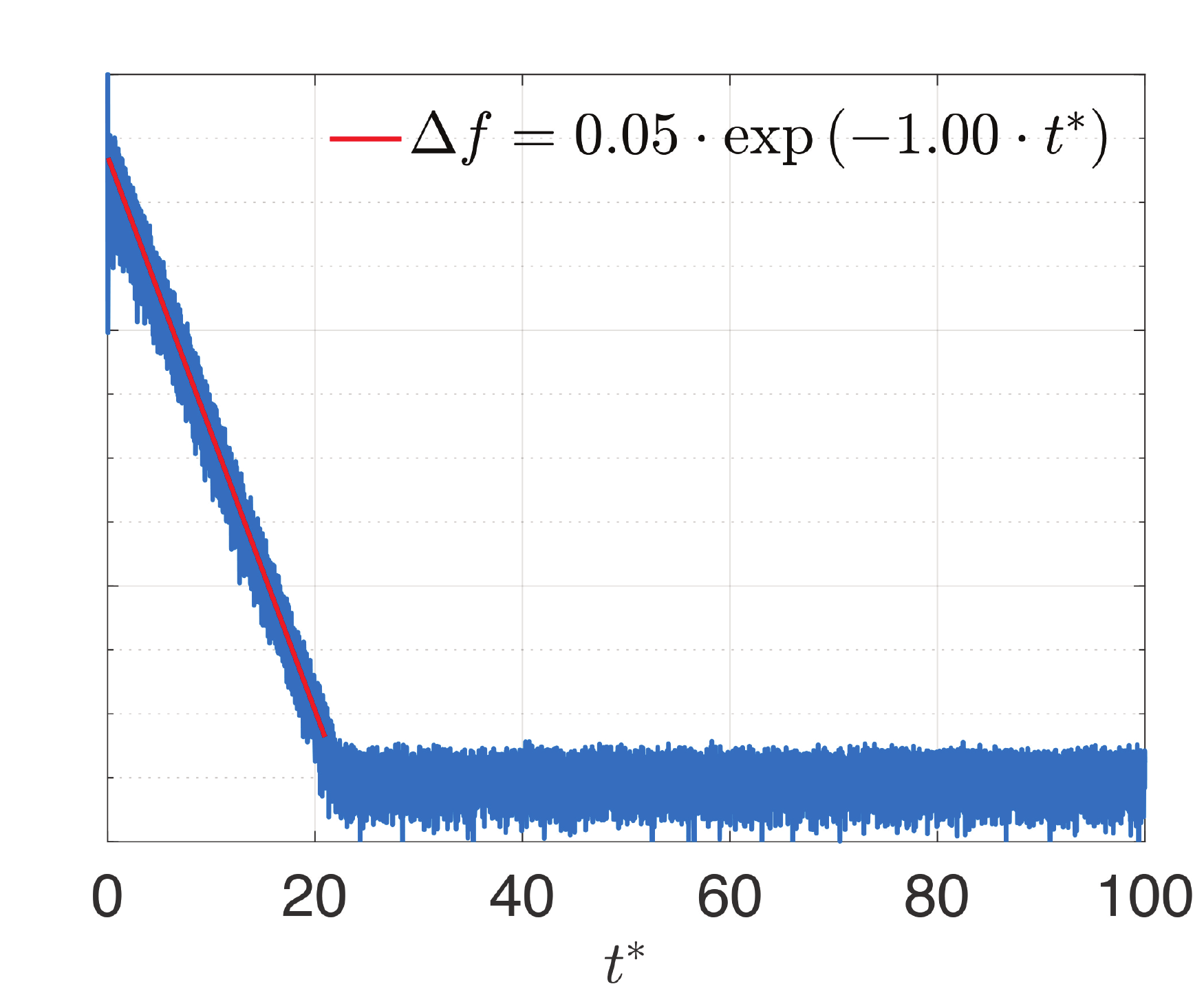}
		\caption{ $ {\sigma = 0.1 } $  and  $ {\Pi_{K} = 10^{5}} $ }
	\end{subfigure}
\begin{subfigure}[b]{0.32\textwidth}
		\includegraphics[height = 0.79\textwidth]{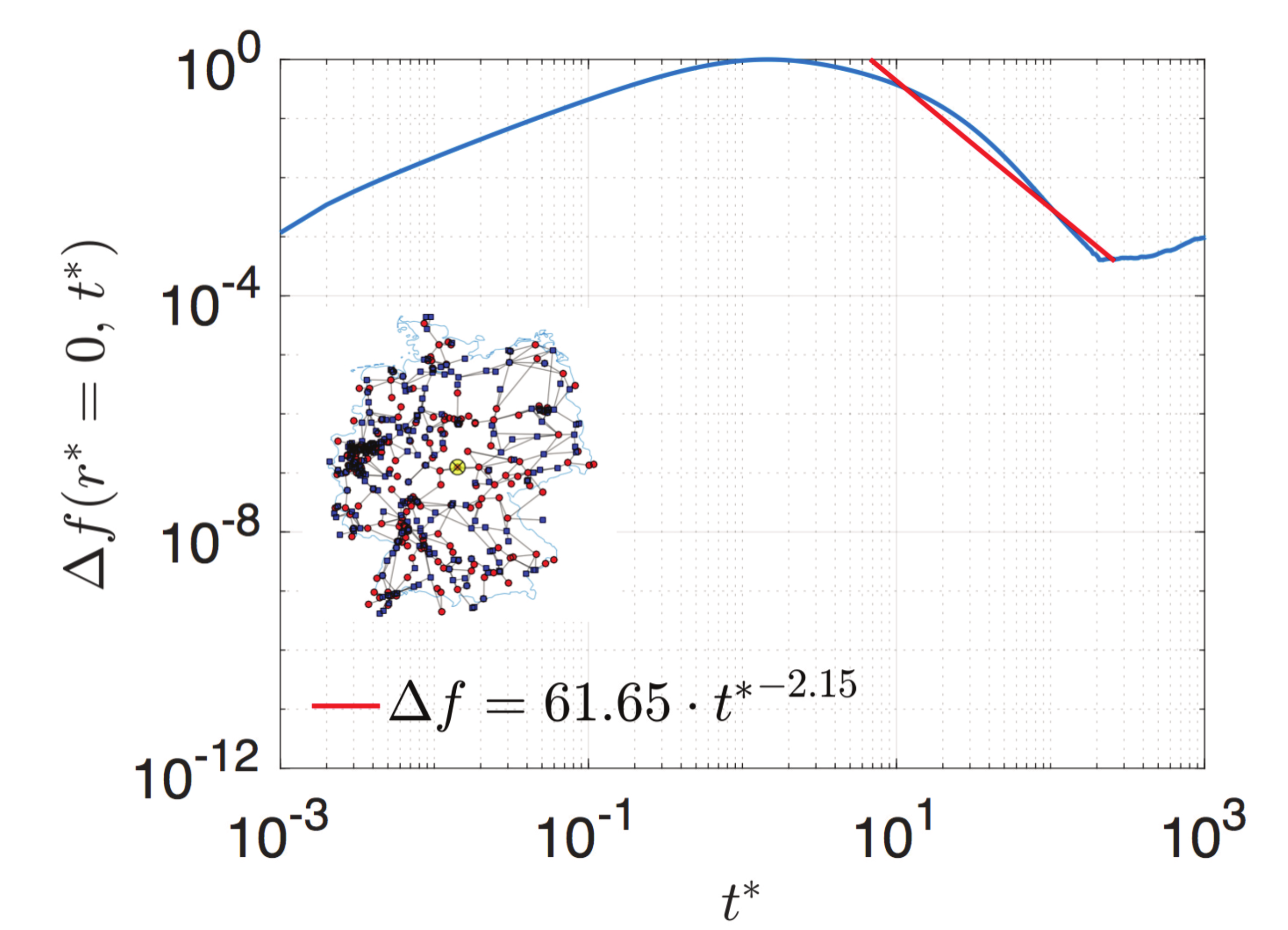}
		\caption{ $ {\sigma = 0.1 } $  and  $ {\Pi_{K} = 10^{-1}} $ }
	\end{subfigure}
	\hspace{.15cm}
	\begin{subfigure}[b]{0.32\textwidth}
		\includegraphics[height = 0.8\textwidth]{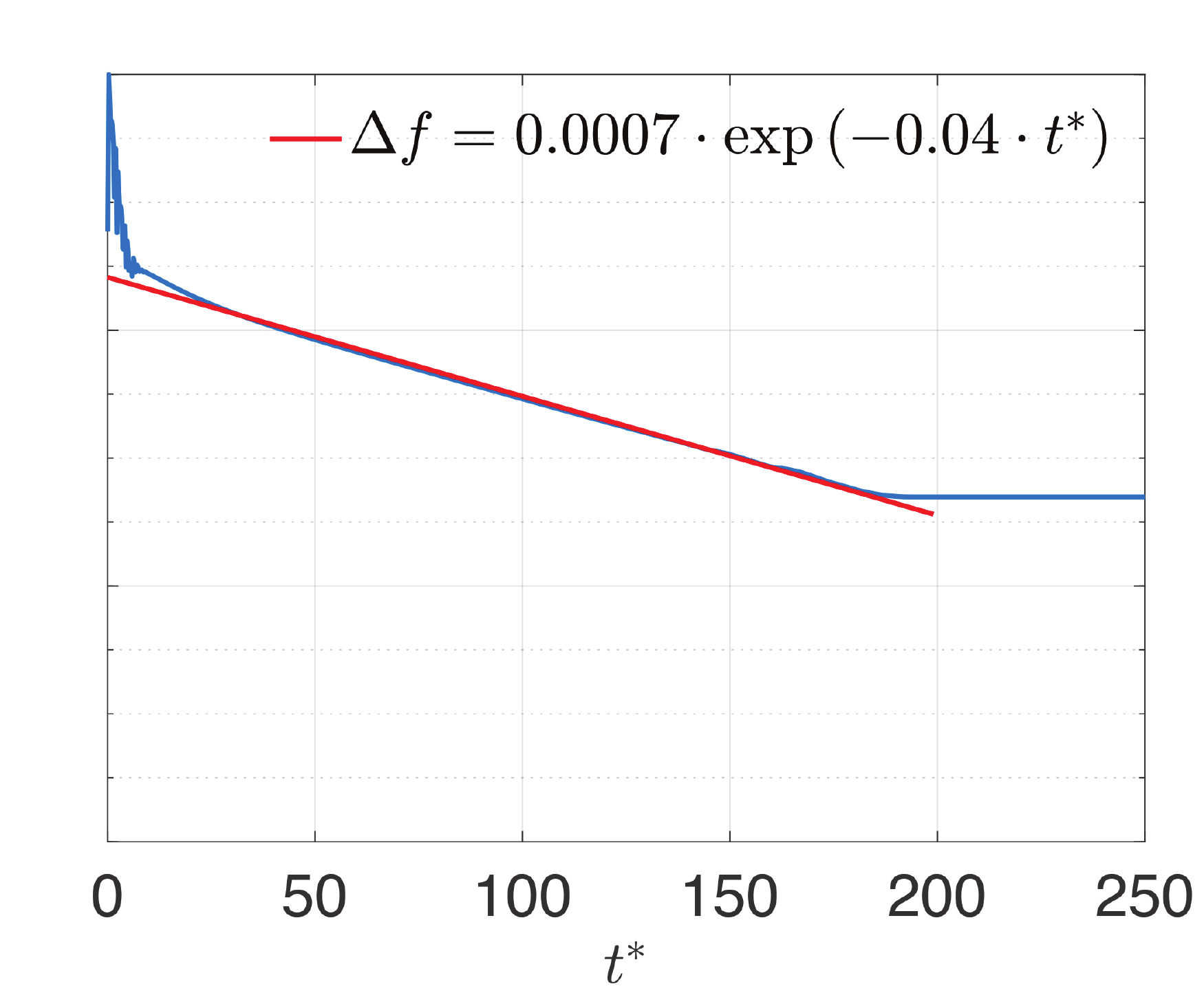}
		\caption{  $ {\sigma = 0.1 } $  and  $ {\Pi_{K} = 10} $  }
	\end{subfigure}
	\begin{subfigure}[b]{0.32\textwidth}
		\includegraphics[height = 0.8\textwidth]{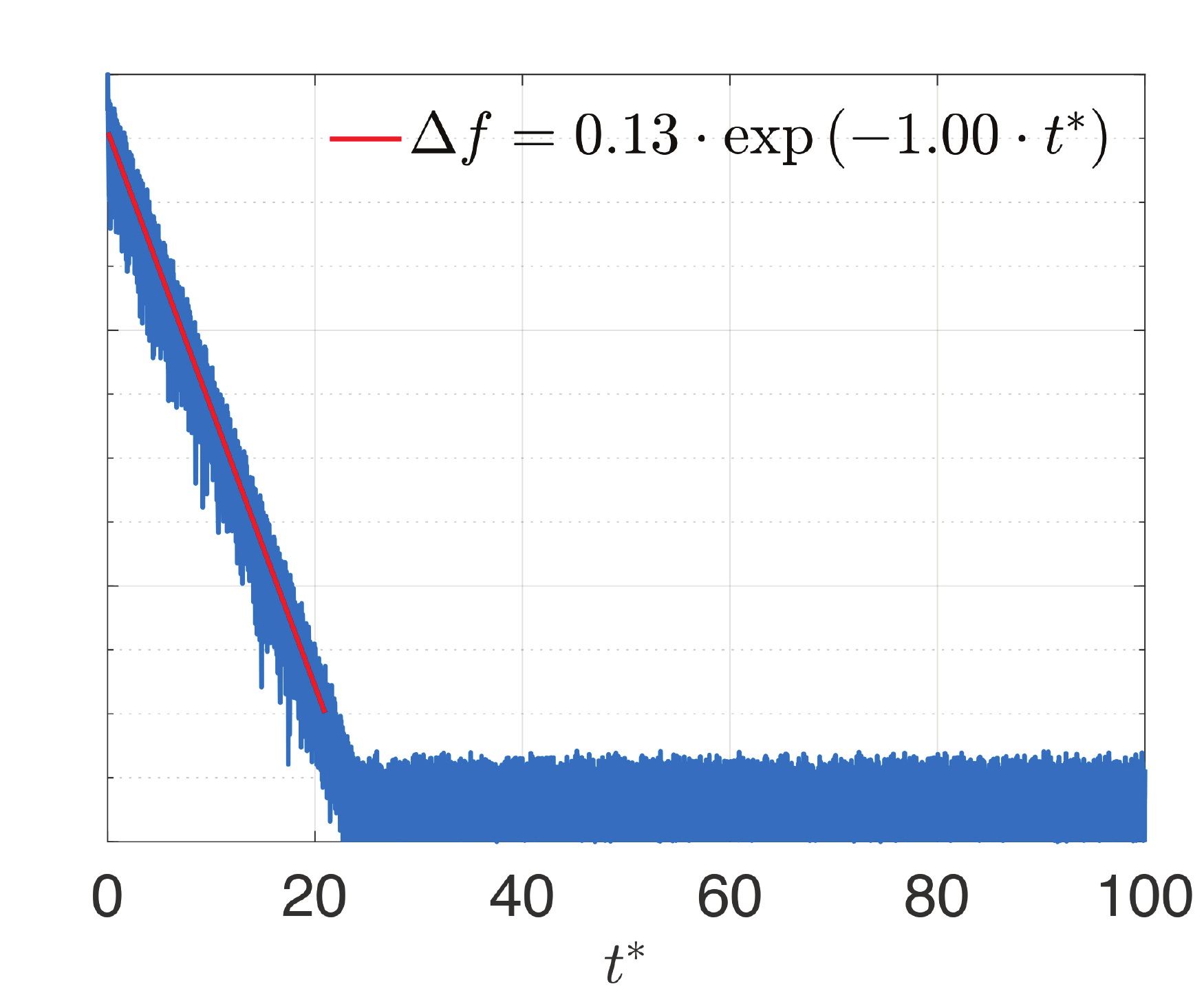}
		\caption{ $ {\sigma = 0.1 } $  and  $ {\Pi_{K} = 10^{5}} $ }
	\end{subfigure}
	\caption{Averaged change of power flow as function of time (Blue Curves) at  $ {r ^{*}= 0} $  with disturbance in power   $ {\delta P = 0.001 P } $ for  
	(a)-(c)  a Cayley tree grid  $ (N=484) $  for  three exemplary sets of parameters,
	(d)-(f)  a square grid,  $ {L = 22, } $  with periodic arrangement, 
	(g)-(i) a square grid,  $ {L = 22, } $  with random arrangement, 
	(j)-(l) German transmission grid with random arrangement of generators and motors.
	 Each for three exemplary sets of parameters correspond to three distinct stability regions.  We fitted  the numerical results (blue) 
	with exponential and power law functions (red).
	}
	\label{transient}
\end{figure*}
   \begin{figure*}
	\centering
	\begin{subfigure}[b]{0.23\textwidth}
		\includegraphics[width = \textwidth]{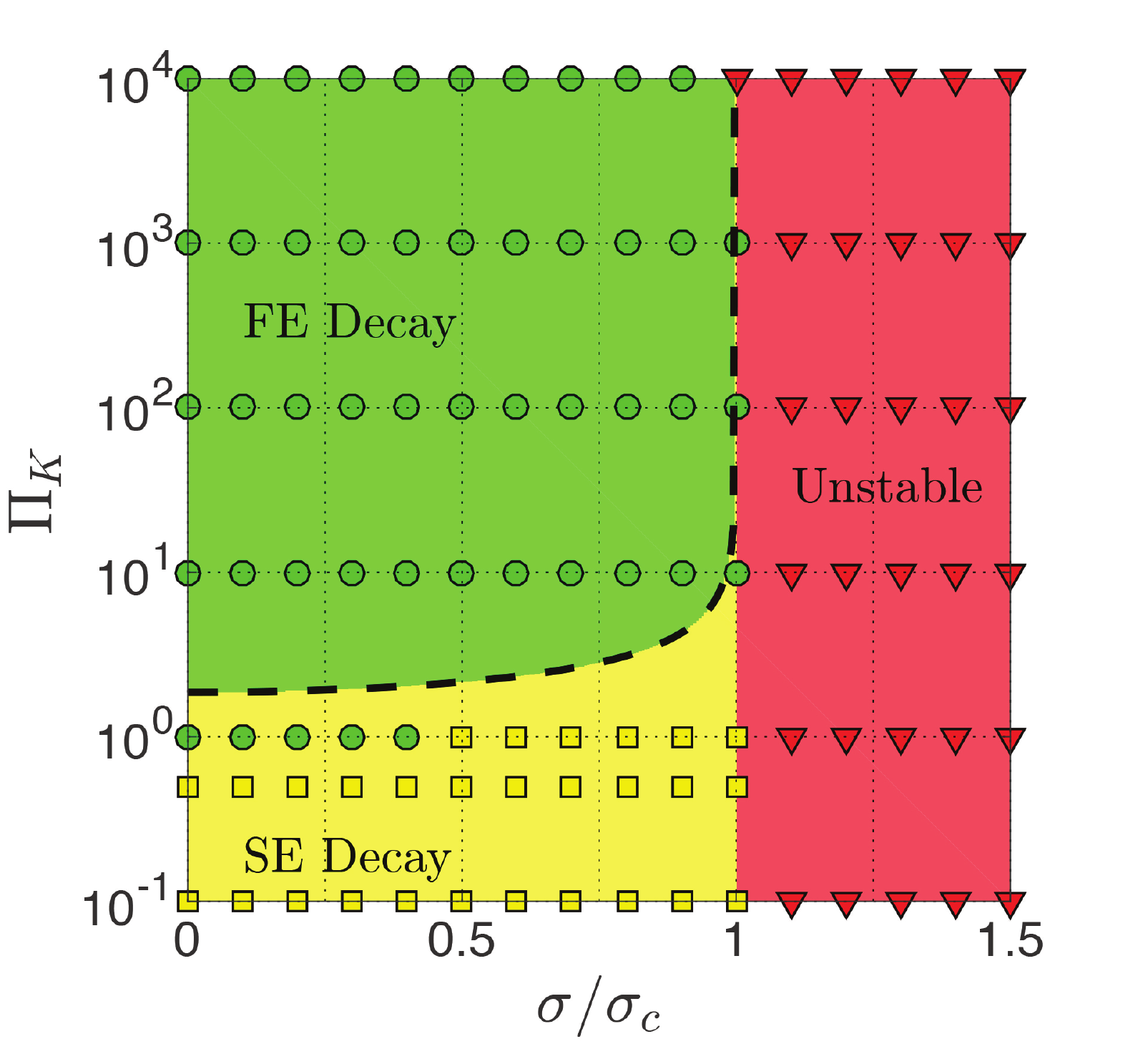}
		\caption{Cayley Tree grid  $ {b=3, } $   \\  $ {N=485, } $   $ {\sigma_{c}=0.20} $ }
	\end{subfigure}
	\hspace{0.1cm}
	\begin{subfigure}[b]{0.23\textwidth}
		\includegraphics[width = \textwidth]{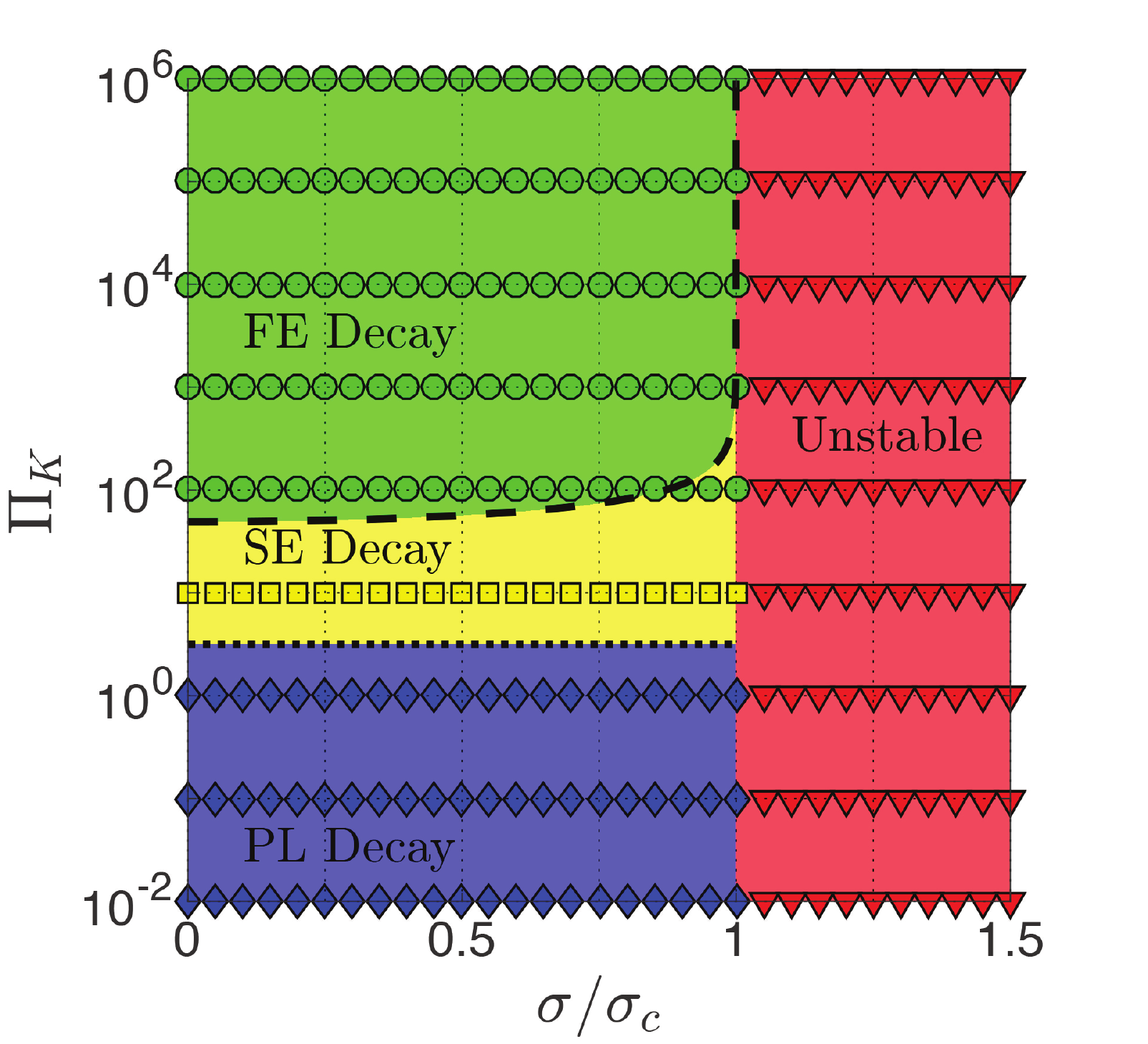}
		\caption{Square grid, periodic   $ {P_i , } $   $ {L=22 , } $   $ {\sigma_{c} = 2.00} $ }
	\end{subfigure}
		\hspace{0.1cm}
	\begin{subfigure}[b]{0.23\textwidth}
		\includegraphics[width = \textwidth]{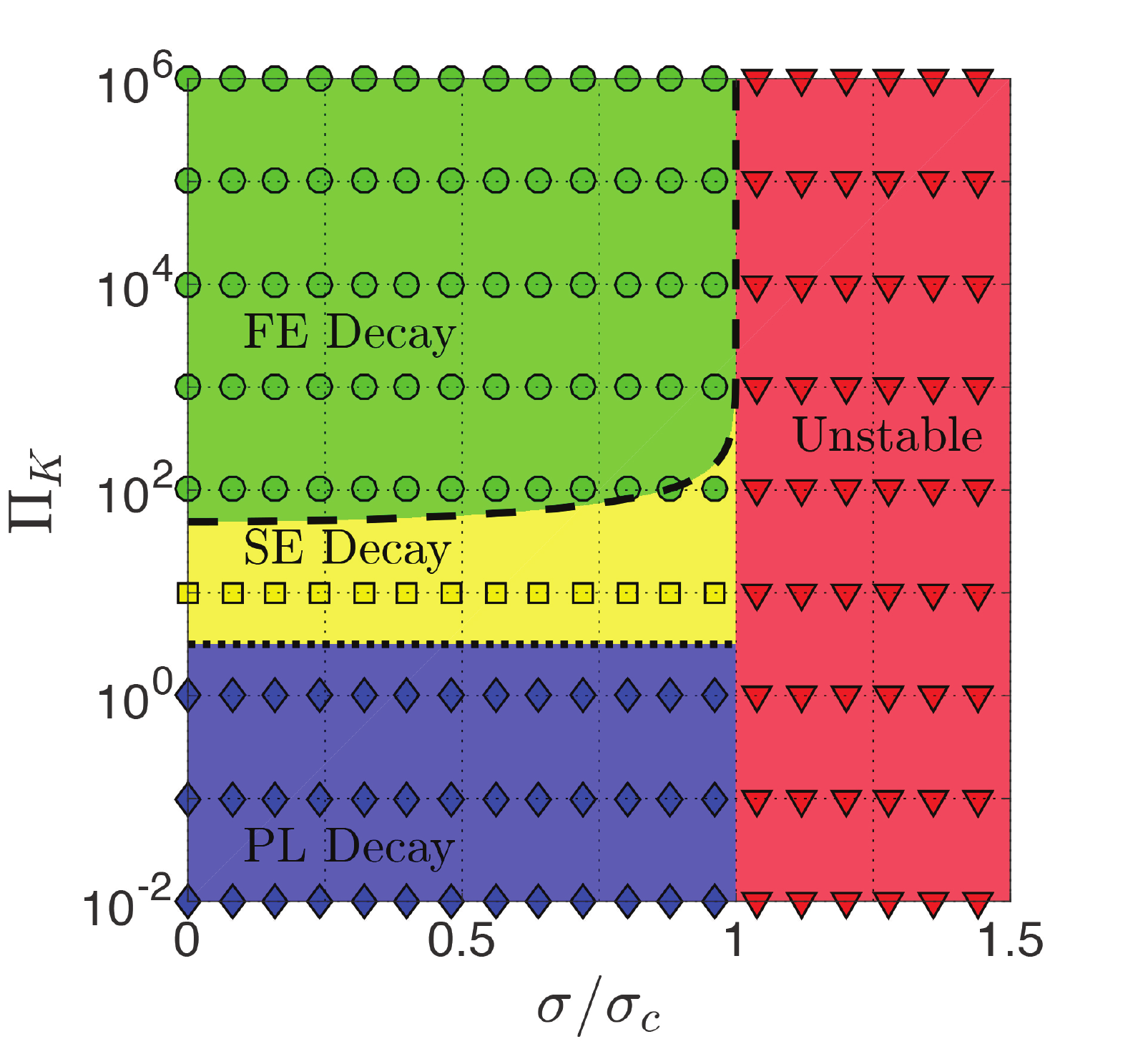}
		\caption{Square grid, random  $ {P_i , } $   $ {L=22 , } $   $ {\sigma_{c} = 1.25} $ }
	\end{subfigure}
		\hspace{0.1cm}
	\begin{subfigure}[b]{0.23\textwidth}
		\includegraphics[width = \textwidth]{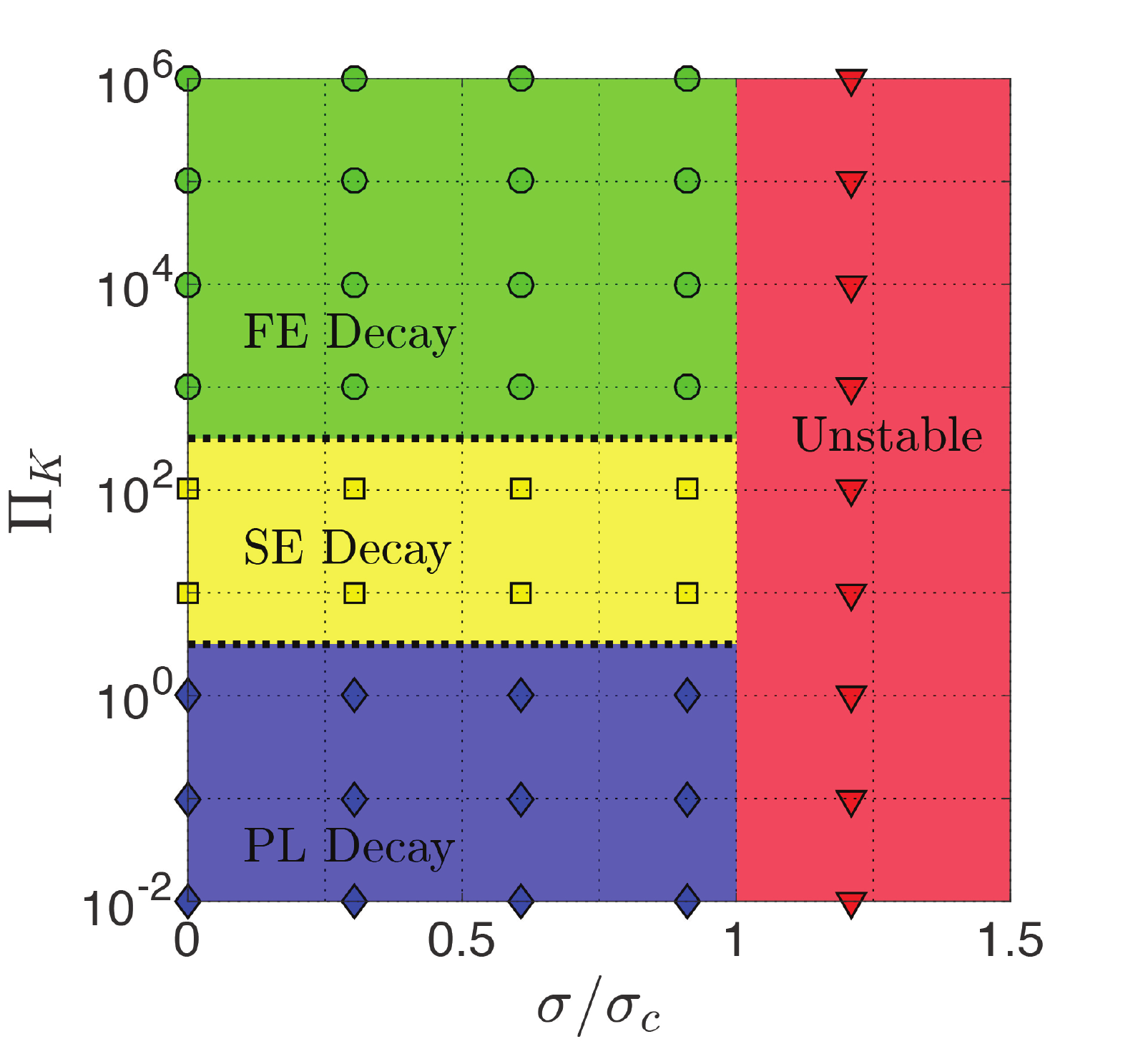}
		\caption{German transmission grid,  $ {N= 502 , } $   $ {\sigma_{c} = 0.33} $ }
	\end{subfigure}
	\caption{Phase diagrams as function of  parameters  $ {\Pi_{K} } $ and  $ {\sigma . } $  Red triangle, green circle, yellow square and blue diamond represent numerically verified parameters that makke the grid unstable, that result in
	fast exponential (FE)  decay,  in slow exponential (SE)
	 decay and  in power law (PL) decay, respectively. 
	Red, green and yellow shaded regions represent  parameters that, 
  according to analytical results,  are unstable, to have FE decay and SE decay,  respectively. Blue shading represents the numerically obtained region  with PL decay. 
	In Fig. (a)
	dashed black line is the phase boundary  between FE and SE
	 decay according to Eq. (\ref{piks}), 
	 in Figs. (b) and (c)  dashed black lines, Eq. (\ref{piksl}). For Figs. (b), (c) and (d), dotted black lines are numerically obtained boundaries.}
		\label{fig:pd}
\end{figure*}

{\bf Classification of Transient Dynamics: Parametric Phase Diagrams.}
 Next, in order to analyze the transient behavior  of disturbances we  
    calculate the absolute values of the  changes in power flow
   in the transmission line 
   between nodes  $ {i } $ and  $ {j, } $  \newline $ {\Delta F_{ij}(r) = | F_{ij}(r) - F_{ij}^{0}(r) | , } $ 
   averaged over all transmission lines at a distance  $ {r } $ from the disturbance and 
    divided  by its maximum value, 
 \begin{align}
{\Delta f}(r^{*},t^{*}) &= \dfrac{\left< \Delta F_{kl}(r^{*},t^{*}) \right>}{\max_t( \langle \Delta F_{kl}(r^{*},t^{*}) \rangle)}.
 \end{align}

In Figs. \ref{transient} we show examples for   transient dynamics
 $ {\Delta f}(r^{*},t^{*})  $ {
 in  three grid topolgies for several different parameter values  $ {\Pi_K,\sigma . } $  If there is no stable solution the 
phase perturbation increases without bound as seen in the example of Fig. \ref{transient} c) for the Cayley tree grid and in the lower phase portrait of Fig. \ref{fig:pp}.
 In order to find   the parametric dependence systematically, 
we varied  both parameters  $ {\Pi_K } $ and   $ {\sigma } $ in small steps.
 By mapping all parameters for which we find  unstable solutions,
we  find   unstable parameter regions for
  $ {\sigma > \sigma_c , } $  as shown 
in red in Fig. \ref{fig:pd} a) for the tree grid, 
in Fig. \ref{fig:pd} b) for the square grid with periodic arrangement, in Fig. \ref{fig:pd} c) for the square grid with random arrangement of consumers and generators, and in Fig. \ref{fig:pd} d) for 
the German transmission grid. The critical values of   $ {\sigma_c } $ depend on the grid
 topologies:
 For the   Cayley tree grid  with  $ {b = 3} $ 
 we find  $ {\sigma_c = 0.2, } $  for the 
  square grid with periodic arrangement    $ {\sigma_c = 2.00, } $  
   for the Square grid with random arrangement    $ {\sigma_c = 1.25} $ 
   and for the German transmission grid   $ {\sigma_c = 0.33. } $ 
In the stable  regions we identify the following 
three qualitatively different transient behaviours: 

The perturbation  decays {\it exponentially fast} (FE) with  local relaxation rate  $ {\Gamma_0 } $ superimposed by oscillations, 
as seen in  exemplary    transients  at the origin of the disturbance  $ {\Delta f(0,t^*) } $ in Fig. \ref{transient} a) for  tree grid, in Fig. \ref{transient} f) for periodic square grid, 
in Fig. \ref{transient} i) for  random square grid and in Fig. \ref{transient} l) for German transmission grid. 
All parameter sets showing FE behavior are plotted 
 in Fig.  \ref{fig:pd} a)-d) as green circles. 

We observe the decay to be exponential with a smaller relaxation rate  $ {\Gamma < \Gamma_0, } $ for a large interval of time  $ {t-t_0 \gg \tau} $ 
 as seen in Figs. \ref{transient} b) for the tree grid, e) and h) for the square grids 
and k) for the German grid. The corresponding parameter 
sets which show such {\it slow exponential relaxation } (SE)  are shown in Fig. \ref{fig:pd} a)-d) as yellow squares.

In the green 
shaded area  fast relaxation (FE) is expected to occur according to 
analytical results of  the next section,  
 limited by  the dashed lines in  Fig. \ref{fig:pd} a)-c),
 a plot of  Eq. (\ref{piks}) for the tree grid and of  Eq. (\ref{piksl}) for the square grids, 
respectively. For the German grid, we indicate the boundary of that region in Fig. \ref{fig:pd} d) by a dotted line, as suggested by numerical 
results. 
    The yellow shaded areas in 
Fig. \ref{fig:pd} a)-c) show where  slow relaxation (SE) is expected to occur according to the analytical results. There  is good agreement, the slight inconsistency  is in a region where we observe deviations of  $ {\Gamma } $  from  $ {\Gamma_0 } $  by only a few 
percent, well within the estimated errorbars of the numerical results.
This parametric dependence of the   relaxation rate 
is also seen in Fig. \ref{fig:gammasigma} for the  $ {b=3} $  Cayley tree grid
  where we plot
  the decay rate  $ {\Gamma(\sigma) } $  in units of local relaxation rate  $ {\Gamma_0, } $  as obtained by fitting the transient behavior with an exponential decay.  
  
\begin{figure}[t!]
        \centering
  \includegraphics[width = 0.75\columnwidth]{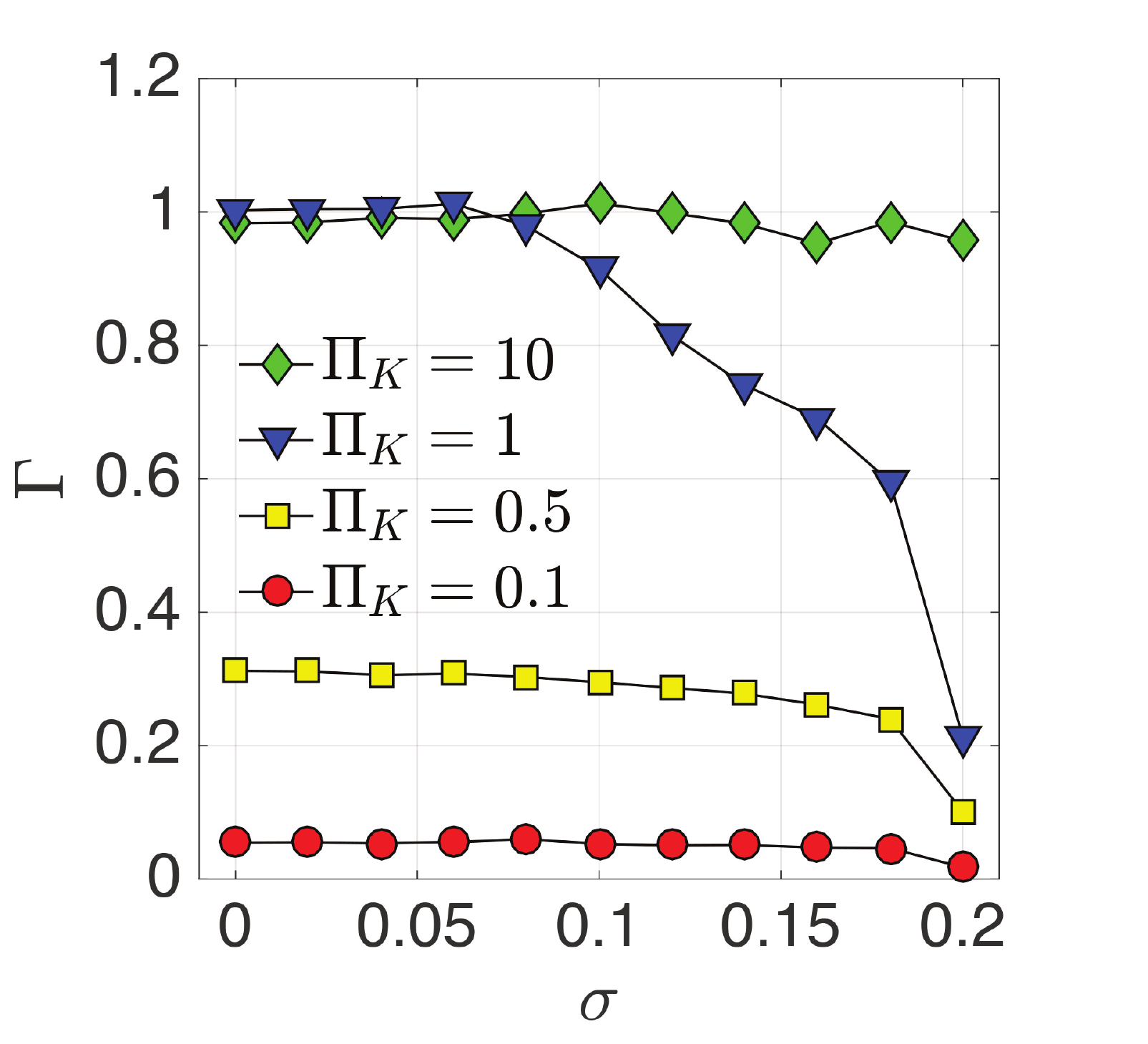}
 \caption{ The decay rate  $ {\Gamma(\sigma) } $ 
  in units of local relaxation rate  $ {\Gamma_0} $ 
 in the  $ {b=3} $  Cayley tree grid for
 different values of  $ {\Pi_K, } $ 
     as obtained by fitting  transient behavior with exponential decay.
       } 
\label{fig:gammasigma}
\end{figure}

Finally, and  most surprisingly, we observe an even slower decay with a {\it power law} in time in  square grids and in the German grid, as seen in  exemplary plots Figs. \ref{transient} d)  and g) for the square grids and j) for the German grid. 
The power is close to  $ {2, } $   in good agreement with  diffusive behavior for  $ { \Delta f(t) , } $  as  
obtained  for a model square grid by an analytical 
 derivation \cite{Kettemann2016}, see the next section,   Eq. (\ref{changeFdynamic}). Although that result was derived  for square grids only, 
the numerical results indicate  that such diffusive behavior may occur
 also in other  meshed grids, as the example of the German grid shows at small values of the parameter  $ {\Pi_K< 2, } $  Fig. \ref{transient} j).
   Thus, although the local relaxation time  $ {\tau = J/\gamma } $ decreases when the inertia  $ {J } $ is lowered, more nodes become correlated and thereby the spreading of a disturbance is slowed down 
to a collective diffusive spreading for times  $ {t>\tau } $ as  $ {\Pi_K } $ and thereby the inertia  $ {J } $ is lowered. 

Diffusive propagation  results also in a slowed spatial 
 spreading  of disturbances, as
  seen by calculating the expectation value of the   squared distance, 
  $ {\langle r_t^2 \rangle = \sum_i \alpha_i^2(t) (r_i-r_j)^2/\sum_i \alpha_i^2(t). } $ 
 Diffusion results 
  in linear 
increase with time for times  $ { t-t_0 > \tau , } $  while ballistic motion  gives a faster,
 quadratic increase.
 In Fig.  \ref{perturbationarea} b)
   the result is shown for a 
    square grid with periodic arrangement of   $ {P_i . } $  
    We find that  $ {\langle r_t^2 \rangle } $  increases initially very fast for  times  $ {t-t_0 < \tau. } $  
   \begin{figure*}[ht!]
	\centering
	\begin{subfigure}[b]{0.24\textwidth}
		\includegraphics[width = \textwidth]{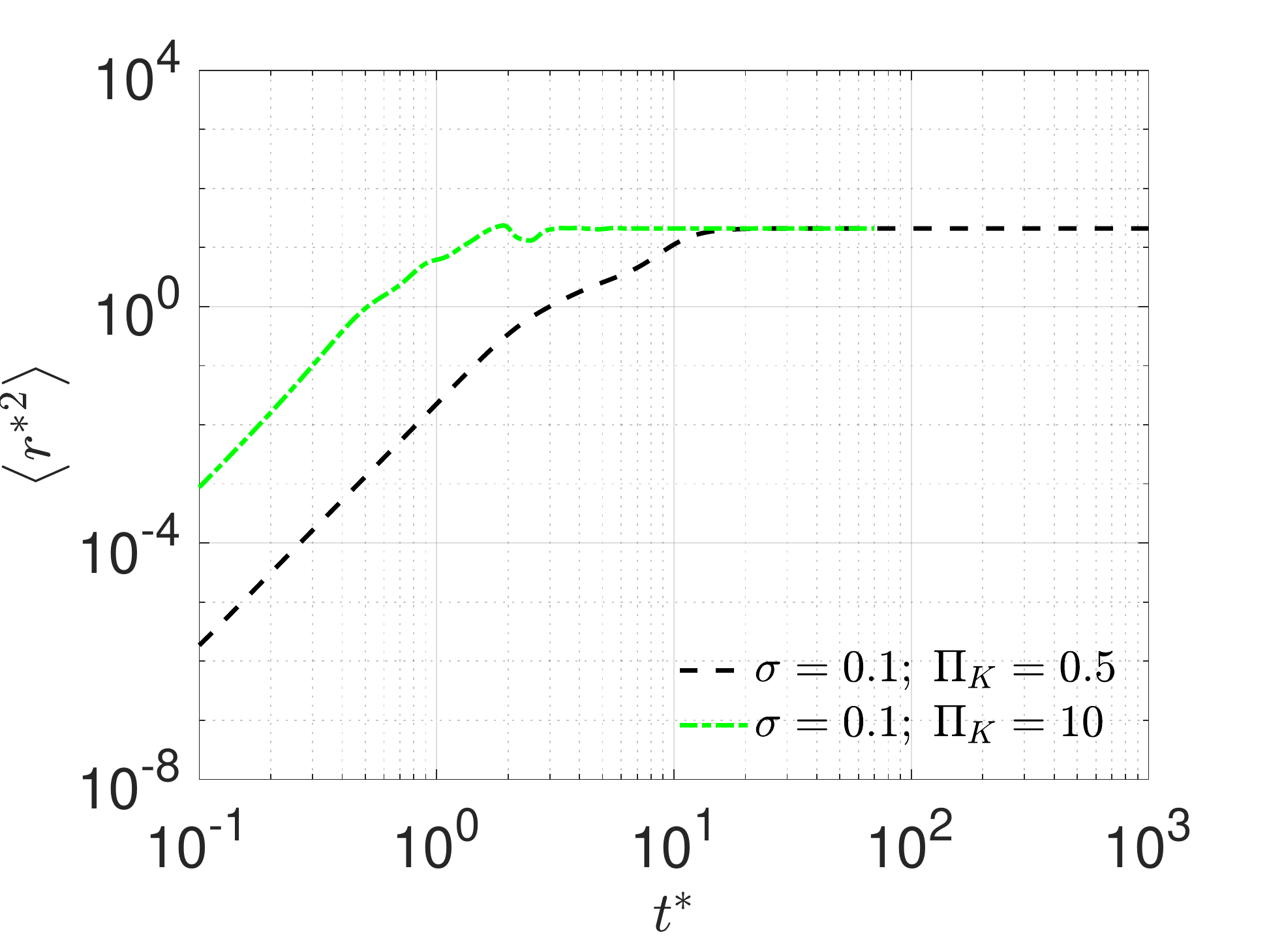}
		\caption{}
	\end{subfigure}
	\begin{subfigure}[b]{0.24\textwidth}
		\includegraphics[width = \textwidth]{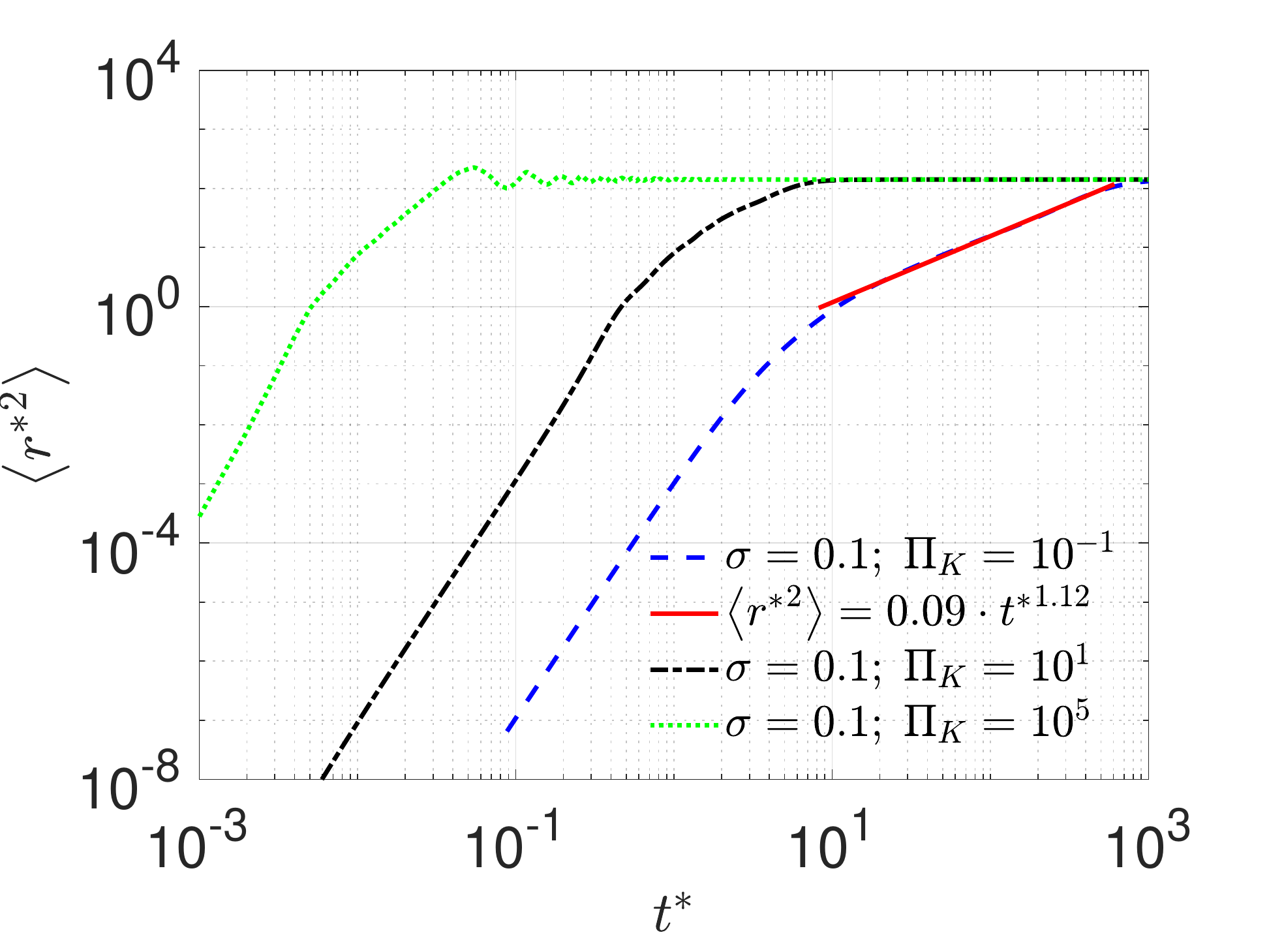}
		\caption{}
	\end{subfigure}
	\begin{subfigure}[b]{0.24\textwidth}
		\includegraphics[width = \textwidth]{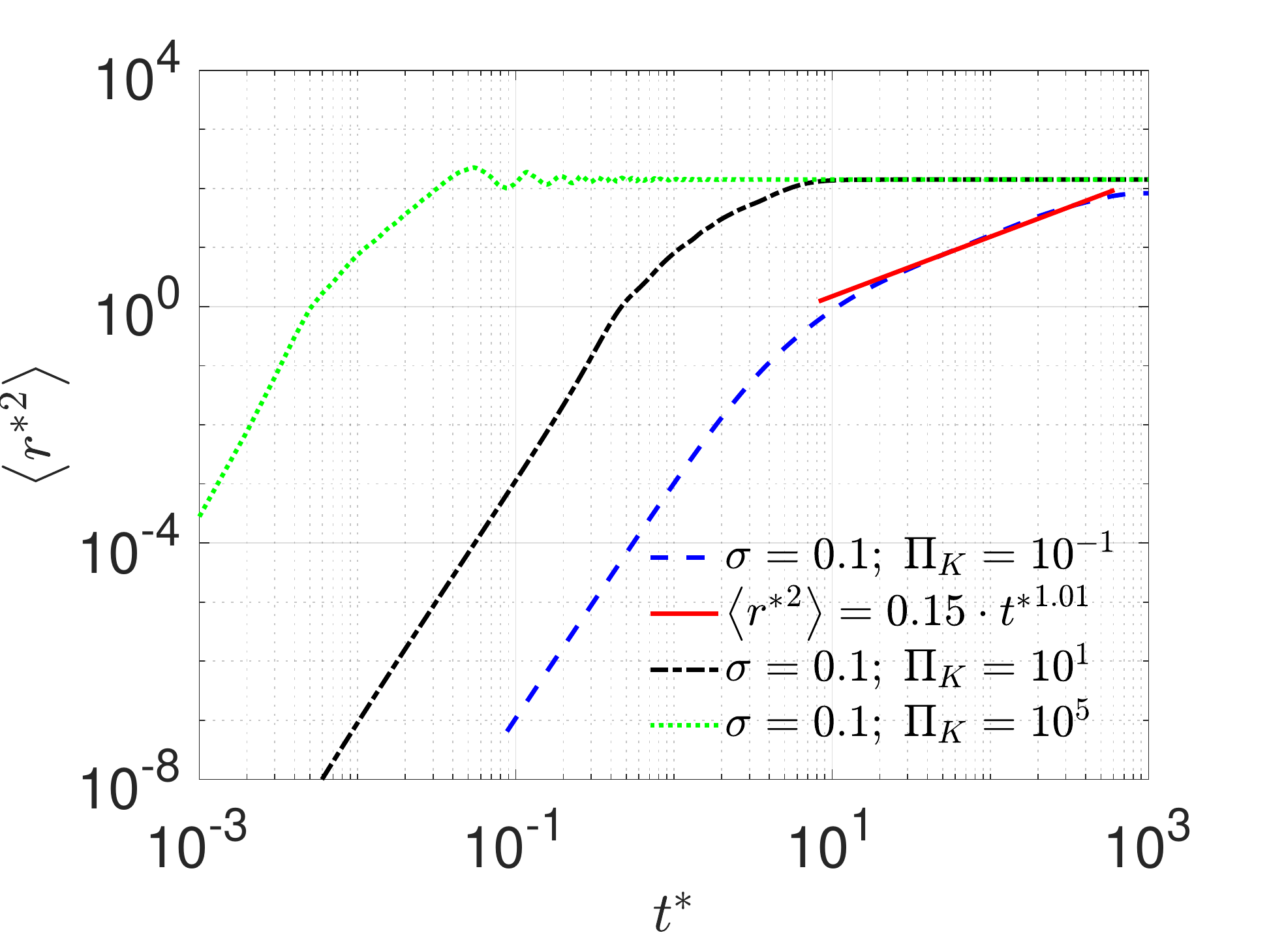}
		\caption{}
	\end{subfigure}
	\begin{subfigure}[b]{0.24\textwidth}
		\includegraphics[width = \textwidth]{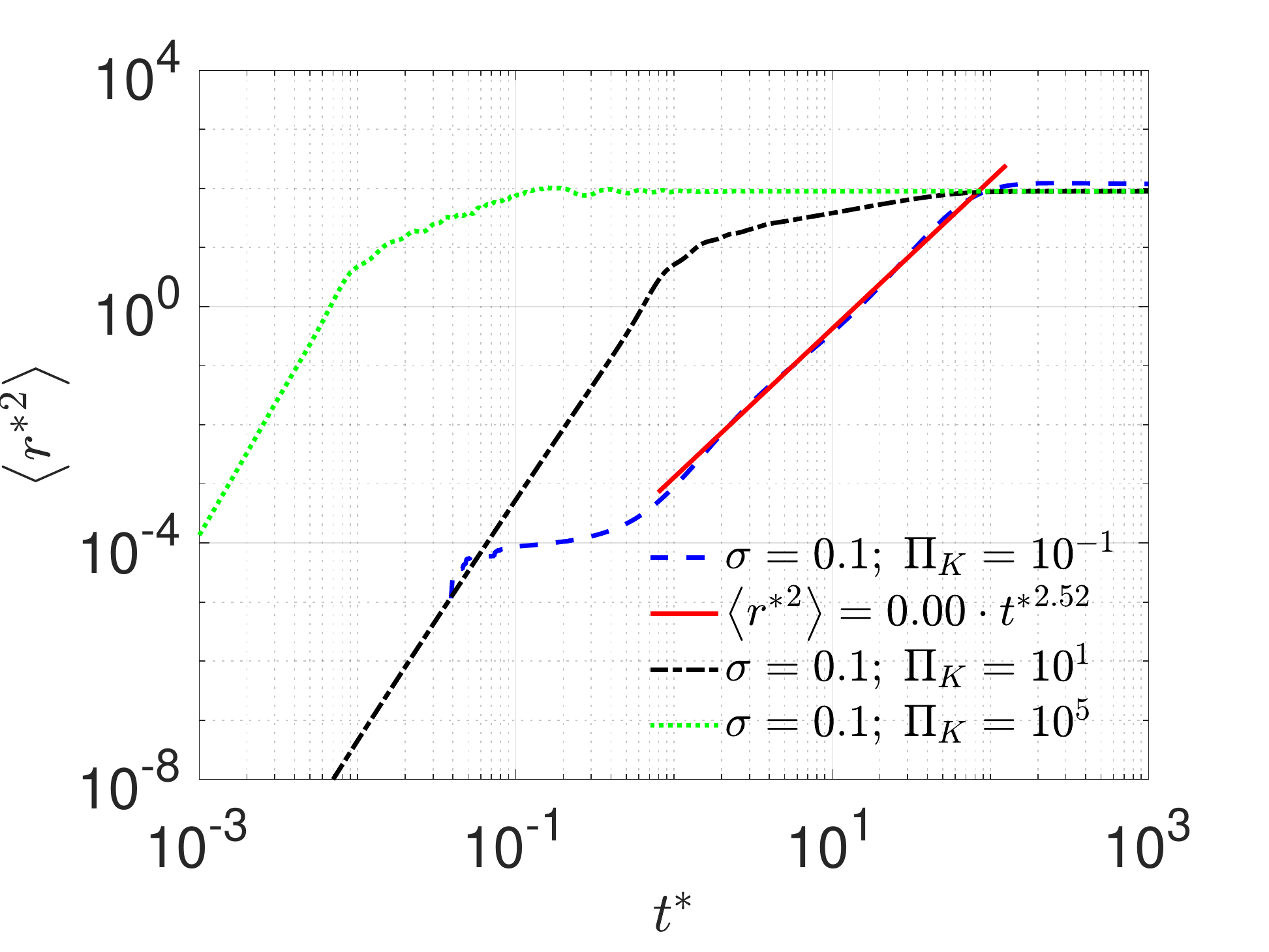}
		\caption{}
	\end{subfigure}
	\caption{Expectation value of the square of the distance  $ {r_i } $ to the origin 
  of  power disturbance 
  $ {\delta P = 0.001 P } $ as function of time, for the exemplary sets of parameters
 given in the insets
	(a) in  a Cayley tree grid  ( $ N=485 $ ),	
	(b) in a square grid ( $ L = 22 $ ) with periodic arrangement, 
	(c) in a square grid ( $ L = 22 $ ) with random arrangement, 
	(d) in the German transmission grid with random arrangement of generators and motors.
	}
	\label{perturbationarea}
\end{figure*}

   For small  $ {\Pi_K=0.1 } $   it  slows down to a power law increase for
      $ {t-t_0 > \tau } $  with 
      $ {\langle r_t^2 \rangle \sim t^{\beta}, } $ 
      where \newline  $ {\beta = 1.12 } $  fits the data, which is a very strong indication 
       of diffusive propagation in square grids with small inertia. Finally,  
        the expectation value of the   squared distance converges 
       to a value of the order of the system area  $ {L^2 } $  for large times  $ {t-t_0 > 1000 \tau . } $  Arranging 
         $ {P_i } $  randomly, we find an even better agreement with diffusive behavior,   $ {\beta =1.01 } $  for 
           $ {\Pi_K=0.1 . } $  On the Cayley tree  grid such a slow increase of  $ {\langle r_t^2 \rangle} $ 
          is absent, it
    reaches a  value of the order of the system area much
    earlier. This is in agreement with our observation  that 
           diffusive power law decay  is absent in the transients of  tree like grids, 
            Fig. \ref{fig:pd} a).  In  the german transmission grid we find for the 
            node marked by a cross    in Fig.  \ref{fig:grid}, that the squared distance $ {\langle r_t^2 \rangle} $ 
 increases more slowly than in the square grid  for $ {\Pi_K=10,}$ see 
          Fig.  \ref{perturbationarea} d), in agreement with the insight that  the spreading is 
           slowed down in meshed 
           grids. 
          At smaller intertia, corresponding to 
            $ {\Pi_K=0.1,}$ it shows for that node 
             a particular behavior with a delayed spreading followed
             by a faster spreading of the disturbance.

 In the next section we present analytical results  for the  transient 
  behavior for model 
 grids. We show that  diffusive spreading of  disturbances,
  is possible
  in  meshed grids like square grids, but that it is absent in tree like grids. This is traced 
 to  the presence of a spectral gap in the  Laplace operator of these grids.

 {\bf  Spectral Analysis of  Disturbances.}
               \begin{figure}[] \label{Fig:spectrum}
        \centering
  \includegraphics[width = 0.75\columnwidth]{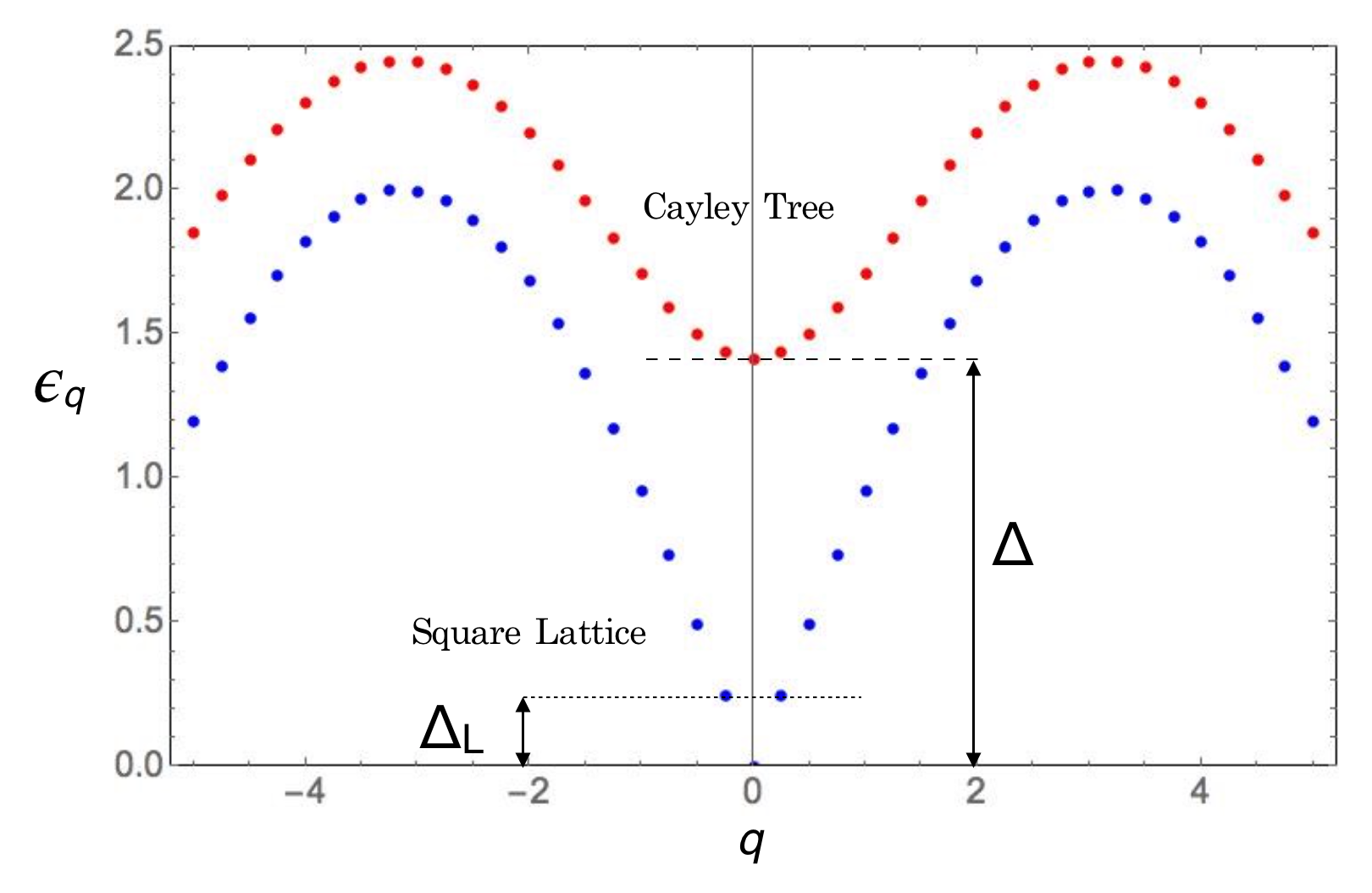}
 \caption{ 
    Dispersion  $ {\epsilon_q } $  as function 
    of discrete wave number  $ {q_n } $ for  
    Cayley tree grid (red dots) and square lattice  (blue dots). 
      The   spectral gap to the first excitation energy
      is  for the Cayley tree grid  $ {\Delta, } $  (dashed line), 
    while for the square lattice the gap
        $ {\Delta_L=\epsilon_{q_1} } $ is smaller (dotted line) and
   decreases with  $ {L . } $ 
} 
 \label{fig:spectrum}
\end{figure}
 {\it Periodic Square Lattice.}  For a square lattice
   where the 
 power capacitance  $ {K } $ is the same for all transmission lines and generator and consumer power  $ {P_i =\pm P } $ is arranged periodically, an analytical solution is obtained
 with the plain wave Ansatz   $ {\phi_{q i} = c_{q}e^{\imath {\bf q r}_i} } $ 
where   $ { {\bf q} } $ is the wave vector.  
   The Eigenfrequency   $ {\epsilon_{q_n}, } $ 
     is obtained  as \cite{Kettemann2016},
 \begin{eqnarray} \label{alphaharmonic2}
   \epsilon_{q_n}  = \sqrt{ \Pi_K}  (1-\sigma^2/\sigma_c^2)^{1/4} \sqrt{ 4 -f_{q_n}} \Gamma_0,
 \end{eqnarray}
 where  
   $ {\sigma/\sigma_c = P/(4 K) } $ and \newline $ {f_{q_n} =2 (\cos q_{n x} a + \cos q_{n y} a) . } $  
   For finite size  $ {L , } $  the wave vectors are quantized,
   $ {q_{x,y n} = n_{x,y} 	\pi/L, } $ 
    with  \newline $ { n_{x,y} = 0, \pm 1,.... }$
     The resulting dispersion of the Eigenfrequency  $ {\epsilon_q } $  is plotted in  Fig. \ref{fig:spectrum} as function 
    of the discrete wave number  $ {q_n } $ (blue dots). 
     We observe    a  gap to the first excitation energy   $ {\epsilon_{q_1} = \Delta_L } $ as
  indicated by the dotted line in 
    Fig. \ref{fig:spectrum}, which decreases with size  $ {L } $ as 
     \begin{equation} \label{gap}
     \Delta_L = \Pi_K^{1/2} (1-\sigma^2/\sigma_c^2)^{1/4} \frac{\pi a}{L} \Gamma_0.
     \end{equation}
   Inserting  $ { \Lambda_n = \tau^2 \epsilon_{q_n}^2 } $ into the Fourier expansion Eq. (\ref{linearresponse}),
     we  get the transient behavior of the phase deviation  $ {\alpha(t) } $ for
      all times  $ {t. } $  
           The condition that slow modes  with  a small relaxation rate
        appear is,  that  the spectral gap  $ {\Delta_L } $ is smaller than the local  
     relaxation rate, 
       $ {\Gamma_0. } $   This yields the 
        parametric condition  $ {\Pi_K < \Pi_K^S(L), } $ 
        where  $ { \Pi_K^S(L)} $ 
        depends on  grid size  $ {L } $  and  ratio 
          $ \sigma $  as
       \begin{equation}  \label{piksl}
       \Pi_K^S(L) =(1-\sigma^2/\sigma_c^2)^{-1/2} \left( \frac{L}{\pi a}\right)^2.
     \end{equation}
This result is  plotted in the phase diagram  Fig. \ref{fig:pd} b) (dashed line)
together with  numerical results.
  If the condition  $ {\Pi_K < \Pi_K^S(L) } $ is fulfilled,    Eigenmodes with  small wave number  $ {q } $   have purely imaginary
    $ {\Omega_q}$ 
   which results in slow decay of the deviation without  oscillations. 
  Summing over all  slow  modes in the spectral representation of  $ {\alpha_i(t) , } $  Eq. (\ref{linearresponse}) 
   we find that  a perturbation at node  $ {j } $ spreads 
 for  times  $ { t> \tau } $ and distances exceeding the mean free path  $ {l = v_0 \tau  } $ 
Thus, the initially localized perturbation 
spreads diffusively with diffusion constant  
 $ {
D =  v_0^2 \tau/2, } $  
   with velocity  $ {v_0 = \sqrt{\Pi_K \cos \delta_0} a/\tau, } $   see Suppl. III for the derivation.
Diffusion causes   slow power law relaxation
     of the disturbance at the initial site and an initial increase, followed by a power law decay at other sites.
          The resulting power law relaxation of the change in   transmitted power between 
            nodes  $ {k } $ and  $ {l } $ is \cite{Kettemann2016}
           \begin{equation} \label{changeFdynamic}
           \delta F_{kl}(t) = \pm  \delta P A_{kl} \frac{\pi^2 a^2  }{\omega_0 D t^2} \exp \left(-\frac{({\bf r_{\rm j} -r_{\rm l}})^2}{4 D t} \right).
           \end{equation}
           Thus, we find analytically that the change in power at the site of the perturbation
             $ {r_{\rm j} } $  decays with a power law in time with power  $ {2 } $  in excellent agreement 
             with the numerical results for  the 
              periodic square grid, Fig. \ref{transient} d).       
The expectation value 
 of the  square of the distance to the perturbation  $ {\langle r_t^2 \rangle, } $ 
  is for the  diffusive propagation at times  $ {t>\tau , } $  Eq. (\ref{changeFdynamic}) obtained to be    $ {\langle r_t^2 \rangle = 2 D t, } $  
  increasing linearly with time  $ {t . } $  
 After the Thouless  time   $ {t_L = L^2/D } $  \cite{thouless}, 
the disturbance reaches the  boundary of the grid
and is reflected.  Then,  for times exceeding  $ {t_L, } $  
the disturbance  decays exponentially with   rate  \newline
   $ \Gamma_{min} = ( 1 -   (1 - \tau^2 \Delta^2)^{1/2} ) \Gamma_0.$ 
%
 
  For square grids with inhomogenous distribution of power  $ {P_i } $ 
 slowly  decaying modes  
   appear when \newline  $ {\Pi_K < \Pi_K^s(L) , } $  with  $ {\Pi_K^s(L) } $  given 
    by Eq. (\ref{piksl}), where  $ {\sigma_c } $ is the critical value 
  for that  distribution of power  $ {P_i . } $   Diffusion  occurs 
  with an accordingly modified diffusion constant  $ {D. } $  
  We plot   $ {\Pi_K^s(L), } $ the dashed line 
  in Fig. \ref{fig:pd} c), together with numerical results  where we use the numerically obtained
   value for  $ {\sigma_c. } $  

{\it Cayley Tree Grid.} 
On a Cayley tree grid every inner node is connected to   $ {d=b+1 } $ other nodes, as shown in Fig. 1 a) for  $ {b=3 . } $    
     For the Cayley
     tree with branching   $ {b>1, } $   the symmetric  eigenvectors 
      were found only recently by Mahan in Ref.  \cite{mahan}.  
 Inserting these Eigenvectors into  the  discrete  wave equations Eq.  (\ref{alpharesponse})
      we   obtain
 the Eigenfrequencies 
       \begin{equation} \label{alphaharmoniccayley}
   \epsilon_{q} = \sqrt{   \Pi_K  \sqrt{1-\frac{\sigma^2}{\sigma_c^2}} \\
(b+1 - 2 \sqrt{b} \cos q a)} \Gamma_0.
 \end{equation}
For $b>1,$ $ {q } $ can not be identified with a wave number since the phase of the 
      Eigenvectors depends nonlinearly on  $ { q a . } $  We plot $ \epsilon_{q}$
       in Fig. \ref{fig:spectrum} for a finite sized tree as function 
    of discrete  number  $ {q_n }.$
        It is remarkable that
  Eigenfrequencies  Eq. (\ref{alphaharmoniccayley})  have for  $ {b >1 } $ a 
    finite gap  $ {\Delta, } $ independent on the number of nodes  $ {N, } $   
    \begin{equation}
    \Delta =  \Pi_K^{1/2}  (1-\sigma^2/\sigma_c^2)^{1/4}
    \sqrt{b+1 - 2 \sqrt{b}} ~ \Gamma _0,
    \end{equation}
    as indicated by the dashed line in 
    Fig. \ref{fig:spectrum}.
      The condition that slow modes  appear is $ { \Delta < \Gamma _0, } $  
    which
      yields the parametric condition  $ {\Pi_K < \Pi_K^s , } $ 
       with 
       \begin{equation} \label{piks}
   \Pi_K^s  =  (1-\sigma^2/\sigma_c^2)^{-1/2}  (b+1 - 2 \sqrt{b})^{-1}.
    \end{equation}
    If that condition is fulfilled, 
     the perturbation decays for large times exponentially 
    with  a reduced relaxation rate 
 } $   
  \Gamma_{min} = ( 1 -   (1 - \tau^2 \Delta(z,\Pi_K)^2)^{1/2} ) \Gamma_0
 $ 
which can be much  smaller than the local relaxation rate  $ {\Gamma_0. } $  
 We plot  $ {\Pi_K^s (\sigma/\sigma_c)} $ 
 in Fig. \ref{fig:pd} a) as the dashed line.
For a  tree  grid with inhomogenous distribution of power  $ {P_i, } $  
   typically, the  slowly  decaying modes  are expected to 
   appear when  $ {\Pi_K < \Pi_K^s , } $  with  $ {\Pi_K^s } $  given 
    by Eq. (\ref{piks}), where  $ {\sigma_c } $ is the critical value 
  for that  distribution of power  $ {P_i. } $


{\bf Discussions.} 
For  the german transmission grid
 the  lines have a typical capacity of  $ {K_{ij} = 10 \text{ GW} } $ \cite{scigridv2}. Assuming that  half of the nodes  act as generators and the other half  as consumers to meet Germany's peak power production of  $ {83 \text{ GW} } $ \cite{agorareport}, we 
set $ |{P_{i}| =  300 \text{ MW} . } $ 
Typical conventional power plants of that rated power  have  inertia $ {J = 10^4 \text{ kg m}^2 } $ and damping  $ {\gamma \omega^2= 0.10 P.} This $ yields $ {J\omega^3 = 310 \text{ GW} }, $   $ {\Pi_{\rm P} = 1.03 \cdot 10^5 } $  
and $ {\Pi_{\rm K} = 3.44 \cdot 10^6. } $  Taking the condition  $ {\Pi_K < \Pi_K^S(L), } $ with Eq. (\ref{piks}) as an estimate that meshed grids show diffusive behaviour, that condition 
becomes for currently existing transmission grids,  $ {L > 1856 a. } $  
Thus, diffusive propagation is unlikely to occur in present transmission grids even on the European scale. However, as conventional power plants become substituted by renewable energy the inertia in the grid  becomes  reduced 
substantially \cite{ulbig}. Thus, the dynamics of transmission grids  changes. For small inertia of  $ {J = 0.1 \text{ kg m}^2 } $ and otherwise same parameters, we find  $ {\Pi_{\rm P} = 1.03} $  and  $ {\Pi_{\rm K} = 34.45 } $ so 
that the condition to observe diffusion is  $ {L > 5.87 a } $ and becomes relevant for transmission grids on the national scale. 
If no measures are undertaken to substitute the inertia of conventional power plants\cite{doerfler}, we conclude that  transient dynamics will change drastically,  disturbances relaxing slowly and spreading by collective diffusive propagation. 

In conclusion, we studied how the relaxation and propagation of disturbances in AC grids is modified when system parameters like the inertia in the grid are changed. To this end we solved the nonlinear swing equations on 
three different grid topologies numerically and analyzed the results comparing them quantitatively with analytical insights obtained by mapping the swing equations for small perturbations on discrete wave equations. 
We solved these wave equations by generalized Fourier expansion for the square grid and the Cayley tree grid. Thereby we show that the long time transient behavior is governed by the spectral gap between the stationary state 
and the lowest Eigenmode of its grid. While the Cayley tree grid has a finite spectral gap, independent on grid size, meshed grids have  a small spectral gap which is reduced strongly with increasing 
grid size, leading to slowed relaxation and collective diffusive propagation of disturbances. 
Analyzing the numerical results we confirm that, depending on inertia, geographical distribution of power, grid power capacity and topology, the disturbance may either decay exponentially in time with the decay 
rate of a single node,  $ {\Gamma^0 , } $  or exponentially with a smaller decay rate  $ {\Gamma < \Gamma^0 , } $  or, even more slowly, decaying with a power law in time, resulting in collective diffusive propagation. Such slow decay of disturbances could be prevented by avoiding meshed grids in favour of tree like grids. 
   
   \section*{Methods}
   
{\bf Stationary solution.}
Before any perturbation is applied, we calculate the stationary state  $ {\theta_i^0} $ 
 at every node  $ {i } $ in the grid. This is accomplished  by first obtaining an analytical solution of Eq. (\ref{Eq:grid_balance})  for small phase differences 
 linearizing   $ {\sin(\theta^0_i-\theta^0_j) \rightarrow \theta^0_i-\theta^0_j . } $  Thereby Eq. (\ref{Eq:grid_balance})
  can be rewritten 
  by introducing  the weighted graph Laplacian matrix   \newline $ {H_{ij}=-K_{ij}+\delta_{ij}\sum_l K_{il} , } $  as 
\begin{equation}\label{Eq:stationary_state}
 P_i=\sum_j K_{ij} (\theta^0_i-\theta^0_j)~~\Rightarrow~~{\bf P}=H\cdot  \mbox{\boldmath $ \theta $ }^0,
\end{equation}
where  $ {{\bf P} } $ and  $ { \mbox{\boldmath $ \theta $ }^0 } $ are vectors, whose  $ {i} $ -th component  
 is the power and  stationary phase at node  $ {i, } $ respectively.
   $ {H } $ has at least one zero Eigenvalue. Therefore,
  we need to calculate the  pseudo inverse  $ {H^{+} , } $ 
   yielding   $ { \mbox{\boldmath $ \theta $ }^0=H^{+} \cdot {\bf P} . } $  
     We use this numerical solution
       then as  initial condition for  a numerical root solver to find the
      solution of the  nonlinear equation, Eq. (\ref{Eq:grid_balance}).
      This way,  the numerical accuracy
      of the stationary solution is maximized which is important, since we 
       use it as   initial condition  for the swing equation to make sure that we 
        start in stationary conditions.
        
%
{\bf Numerical Solution of Swing Equations.} 
Having found the  stationary  phases  $ {\theta_i^0 } $ as solutions of the stationary state equations Eq.  (\ref{Eq:grid_balance}),
we can  insert them   into the swing equations Eq. (\ref{Eq:alpha_norm}). Next, we  solve these equations to study the transient bahavior 
when the AC grid is perturbed by a local disturbance as outlined in detail in Suppl. I. 
We study 
the transient phase deviation  $ {\alpha_i } $ and  the phase velocity  $ {\partial_t \alpha_i, } $ 
 as well as the change in transmitted power
 between all connected nodes  $ {i,j , } $   $ {\Delta F_{i,j} . } $ 
 In this article, we   are primarily  interested
  to understand  the propagation of disturbances  which are so small
   that they do not destabilize the grid.

{\bf Analytical Derivation.} 
                  For small perturbations,
corresponding to the parametric condition  $ { \sigma  < \sigma^* (\alpha), } $  we can
 analyse 
 the swing equations  Eq. (\ref{Eq:alpha_norm}) by expanding it 
           in the perturbation  $ {\alpha_i -\alpha_j . } $  This
        yields the  linear wave equations on the grid \cite{Kettemann2016}, 
                      \begin{equation} \label{alpharesponse}
                 \tau^2  \partial_t^2   \alpha_i  + 
                   2 \tau  \partial_t   \alpha_i 
                   +
       \sum_j   t_{ij}    (\alpha_i -\alpha_j) \nonumber = \delta \Pi_i (t),   
          \end{equation}
            with  coupling amplitudes  $ {t_{ij} =  \Pi_{K ij} \cos(\theta^0_i - \theta^0_j) , } $ 
          depending both on   grid topology and on 
            the initial distribution of power  $ {P_i } $ through the stationary phases  $ {\theta^0_i . } $ 
            These coupling amplitudes  $ {t_{ij} } $ 
           form    the  generalized Laplace operator 
            $ {\Lambda } $ with  $ {\Lambda_{ij} = - t_{ij} } $ and   \newline  $ {\Lambda_{ii} = \sum_i t_{ii}, } $ 
           which is related to  the stability  matrix in the linear stability analysis \cite{coletta,motter}.
The dimensionless disturbance  function is defined by \newline 
        $ {\delta \Pi_i (t) = J \delta P_i(t)/(\gamma^2 \omega). } $ 
 We expand the phase deviation  $ {\alpha_i (t) } $ and the disturbance  $ { \delta \Pi_i (t)} $ 
  in a generalized Fourier series  in terms of the Eigenvectors  $ { {\bf \phi}_n } $  of
          $ {\Lambda, } $  as defined by  
            $ {\Lambda {\bf \phi}_n = \Lambda_n {\bf \phi}_n , } $  where
             $ {\Lambda_n } $  is its Eigenvalue \cite{Kettemann2016,torres,motter,coletta}, see Suppl. III for a detailed derivation. 
              For a local perturbation at a site  $ {j, } $ lasting only a short time interval
           $ {\Delta t \ll \tau } $  around time  $ {t_0 , } $   
            $ {\delta \Pi_i (t) = \delta \Pi \delta_{i j} \tau \delta (t-t_0), } $         
        we find
           \begin{equation} \label{linearresponse}
           \begin{split}
                  \alpha_i(t>t_0) &= - \frac{\delta \Pi}{2} \sum_n  \phi_{n i} \phi_{n j}^* \frac{1}{\sqrt{1-\Lambda_n}} \\
  &\quad \times \left(  e^{-\i \Omega_{n +} (t-t_0)} - e^{-\i \Omega_{n -} (t-t_0)} \right),
  	\end{split}
              \end{equation} 
              where  $ {\Omega_{n \pm} = - \i  (1 \pm \sqrt{1-\Lambda_n}) 1/\tau. } $ 
       In order to find the transient dynamics for all times  $ {t , } $  
       it remains to find the Eigenvalues  $ {\Lambda_n \epsilon_n^2 \tau^2 } $ and Eigenvector components  $ {\phi_{n i} , } $  as can be done numerically for arbitrary grids. For particular grids, analytical solutions can be obtained as we reviewed above and 
        in  Suppl. III.

\section*{Acknowledgements}
We gratefully acknowledge the support of BMBF in the frame of CoNDyNet FK. 03SF0472D. 
       
              \section*{Supplementary} 
     
     More  information on methods and derivations  is provided in the appended supplementary text. For further illustration 
     we also provide three exemplary movies showing the propagation of disturbances
      in a tree grid, a square grid and the german transmission grid. 
       
       \section*{Author contributions} 
       Research design and numerics were performed by S.T and M.C., with S.K. supervising the project and performing the analytical calculations. All authors contributed to editing the manuscript.
      
         \section*{Competing financial interests} 
       The authors declare no competing financial interests.
             
\bibliographystyle{elsarticle-num}

\section*{References}
\begin{thebibliography}{40}

\bibitem{ulbig}
Ulbig, A., Borsche, T. S. \& Andersson, G. Impact of low rotational inertia on power system stability and operation. {\it{IFAC Proceedings Volumes}} {\bf{14}}(3), 7290-7297 (2014).

\bibitem{Kundur1994} 
Kundur, P. Power System Stability and Control. (Mc Graw Hill, 1994). 

\bibitem{Machowski2008}
Machowski, J., Bialek, J.W. \& Bumby, J.R. Power System Dynamics: Stability and Control. (Wiley, 2008). 

\bibitem{Bergen1981} 
Bergen, A.R. \& Hill, D.J. A structure preserving model for power system stability analysis. {\it{IEEE Trans. on Power App. and Syst.}} {\bf{100}}(1), 25-35 (1981). 

\bibitem{Filatrella2008}  
Filatrella, G., Nielsen, A.H. \& Pedersen, N.F. Analysis of a power grid using a Kuramoto-like model. {\it{Eur Phys. J. B}} {\bf{61}}(4), 485-491 (2008).

\bibitem{Rohden2012}
Rohden, M., Sorge, A., Timme, M. \& Witthaut, D. Self-organized synchronization in decentralized power grids. {\it{Phys. Rev. Lett.}} {\bf{109}}, 064101 (2012).

\bibitem{Schmietendorf2014} 
Schmietendorf, K., Peinke, J., Friedrich, R. \& Kamps, R.O. Self-organized synchronization and voltage stability in networks of synchronous machines. {\it{Eur. Phys. J. Spec. Top.}} {\bf{223}}(12), 2577-2592 (2014). 

\bibitem{Kettemann2016} 
Kettemann, S. Delocalization of disturbances and the stability of AC electricity grids. {\it{Phys. Rev. E}} {\bf{94}}, 062311 (2016). 

\bibitem{scigridv2}
Matke, C., Medjroubi, W. \& Kleinhans, D. {\it SciGRID - An Open Source Reference Model for the European Transmission Network (v0.2)}, (July 2016).


\bibitem{Heuck2013} 
Heuck, K., Dettmann, K.D. \& Schulz, D. Elekrtische Energieversorgung, 9th edition. (Springer, 2009). 
 
\bibitem{Rohden2014}
Rohden, M. Self-organized Synchronization in Decentralized Power Grids. PhD thesis, (G\"ottingen, 2014).
 
\bibitem{Menck2014} 
Menck, P.J., Heitzig, J., Kurths, J., \& Schellnhuber, H.J. How dead ends undermine power grid stability. {\it{Nature Communications}} {\bf{5}}, 3969 (2014).

\bibitem{Newman}
M. Newman, Networks: An Introduction. (OUP, Oxford, 2009).

\bibitem{coletta}
Coletta, T. \& Jacquod, P. Linear stability and the Braess paradox in coupled-oscillator networks and electric power grids. {\it{Phys Rev E}} {\bf 93}, 032222 (2016).

\bibitem{motter}
Nishikawa, T. \& Motter, A.E. Comparative analysis of existing models for power grid synchronization. {\it{New J. Phys.}} {\bf 17}, 015012 (2015).

\bibitem{torres} L. A. Torres-S\'anchez, G. T. Freitas de Abreu, S. Kettemann,
Analysis of the Dynamics and Topology
Dependencies of Small Perturbations in Electric
Transmission Grids,
 subm. to 
 IEEE Power Systems,  arXiv:1706.10130 (2017). 

\bibitem{thouless} 
Edwards, J.T. \& Thouless, D.J. Numerical studies of localization in disordered systems. {\it{J. Phys. C: Solid State Phys.}} {\bf{5}}, 807 (1972).

\bibitem{mahan} 
Mahan, G.D. Energy bands of the Bethe lattice. {\it{Phys. Rev. B}} {\bf 63}, 155110 (2001). 
  
%
%
%
%
%
%
%
%
%

\bibitem{agorareport}
Bayer, E. Report on the German Power System, {\it{Country Profile 057/03-CP-2014/EN, Agora Energiewende}}. (Berlin, Germany, 2015).

\bibitem{doerfler} T. Borsche and F. D\"orfler, On Placement of Synthetic Inertia with Explicit Time-Domain Constraints, subm to  IEEE Transactions on Power Systems, https://arxiv.org/abs/1705.03244 (2017).

 
\bibitem{RK} 
E. Kreyszig, Advanced Engineering Mathematics, 9th edition., Wiley, New York (2006).


\end{thebibliography}

\section*{Supplementary} 
\begin{appendix}
\renewcommand*{\thesection}{\Roman{section}}
\renewcommand*{\theequation}{\arabic{equation}}

  \section{Numerical Simulations}

We employ a standard differential equation solver,  
the Runge-Kutta algorithm \cite{RK} using the commercial software MATLAB\textsuperscript\textregistered. 
Since this is a shooting method its convergence is improved considerably by setting the phases to the stationary state solutions before the perturbation. 
The perturbation is applied at $t=0$. Calculated time spans prior and past perturbation were iteratively adjusted to ensure both a stationary state onto which the perturbation is applied and to capture the whole perturbation event until complete decay.
Temporal resolution was chosen fine enough to avoid undersampling of the oscillating phases using the phase portraits $\dot{\alpha}_i(\alpha_i)$ as sensors. They would show smooth curves for sufficient resolution or angled curves for too poor resolution.
In favour of automatized calculation, time span and resolution were mostly not adapted but rather chosen better than necessary. Concretely, for German and square grid, throughout all values of $\Pi_K$, $t \in [-1000\tau, 1000\tau]$ in steps of $10^{-3}\tau$.
For the Cayley tree grid, time span and resolution were i) $t \in [-1000\tau, 1000\tau]$ in steps of $10^{-3}\tau$ for $\Pi_K<10$ and ii) $t \in [-30\tau, 70\tau]$  in steps of $10^{-3}\tau$ for $\Pi_K\geq10$.

\section{Stability}

Depending on the magnitude of the disturbance it can destabilize the grid already at 
smaller values of $\sigma$ than the critical value $\sigma_c$ above which there
 is no stationary solution, 
$\sigma < \sigma_c$. In order to get a typical upper limit for the size of the perturbation $\alpha$ before 
it kicks the system out of the stability, let us first disregard the dependence of the phase deviation at node $i$, $\alpha_i,$ on the perturbation at neighbored sites $\alpha_j.$ This reduces the swing equations
Eq. M(5) of the main article to the one of a single damped, driven nonlinear pendulum. For large times $t \gg 0$ it is well known to have two stable solutions: {\it 1. The stationary solution:} There 
is a stable fixed point at $\partial_t \alpha_i =0$, $\alpha_i = n 2 \pi,$ $n$ integer, to which small deviations relax exponentially fast with the local decay rate $\Gamma_0 = 1/\tau$. {\it 2. The 
over-swinging pendulum solution:} when the driving force and damping are in balance, the phase velocity oscillates around the value $\delta \omega_i = P_i/(2 \gamma \omega)$. 
There are saddle point solutions at $(\alpha_{si},\partial_t \alpha =0)$, where $\alpha_{si}$ is given by 
\begin{equation} \label{saddle}
  \alpha_{s i} = - 2 {\rm arctan \left( \frac{\sum_j \Pi_{K ij} \cos (\theta^0_i-\theta^0_j)}{\sum_j \Pi_{K ij} \sin (\theta^0_i-\theta^0_j)} \right)}.
\end{equation} 
The condition for phase points to lie inside the stability region at node $i$ is then obtained from this local stability analysis to be approximately given by \cite{Kettemann2016}
\begin{equation} \label{basinofattraction}
  \alpha_i^2 + \frac{(\tau \partial_t \alpha_i)^2}{  (1+\sqrt{1-\sum_j \Pi_{K ij} \cos (\theta^0_i-\theta^0_j-\alpha_{si})})} \ll \alpha_{si}^2.
\end{equation}
Thus, we can ensure stability against the perturbation $\alpha (t),$ by making sure that it satisfies the stability condition Eq. (\ref{basinofattraction}) for all times $t$. While that depends on the power 
distribution $P_i$ and the topology of the grid through the stationary phase angles $\theta^0_i$, we can get a typical upper limit for the allowed size of the perturbation $\alpha$ by substituting $\sin (\theta^0_i-\theta^0_j)$ with 
the typical value of $P_i/(d_i K)$, which we denoted above by $\sigma/\sigma_c$.
Substitution into Eq. (\ref{saddle}) gives the saddle point value  \newline $\alpha_s = - 2 { \arcsin} (\sqrt{1-  \sigma^2/\sigma_c^2}).$ Thus, for fixed perturbation amplitude $\alpha$ we find a critical value 
of $\sigma$, above which the disturbance causes instability,
\begin{equation}
   \sigma^* ( \alpha) =  \sigma_c \cos (\alpha/2).
\end{equation} 
We see that the disturbance can destabilize the grid at smaller values $ \sigma^* (\alpha) < \sigma_c$. $ \sigma^* (\alpha)$ coincides with  $\sigma_c$ only in the limit, when the perturbation amplitude is 
vanishing, $ \sigma^* (\alpha \rightarrow 0) = \sigma_c$.
                
  \section{Response to Disturbances}
           
          Stating from the    
        linearized wave equation in the presence of a disturbance, obatined by a 
        linear expansion in phase perturbation $\alpha_i$ in the presence of a fluctuation in power $\delta P$,
\begin{equation}
 \tau^2\partial_t\alpha_i + 2\tau\partial_t\alpha_i = \sum_j t_{ij}(\alpha_i-\alpha_j)+\frac{\delta P_i(t)}{J\omega}\tau^2
\end{equation}
we define the weighted Laplacian $\Lambda$ with
\begin{equation}
 \Lambda_{ij}=-t_{ij}~~\text{and}~~\Lambda_{ii}=\sum_j t_{ij}
\end{equation}
to obtain
\begin{equation}
 \tau^2\partial_t\vec{\alpha} + 2\tau\partial_t\vec{\alpha} + \Lambda\vec{\alpha} =\frac{\delta P_i(t)}{J\omega}\frac{J^2}{\gamma^2}~~.
\end{equation}
Since $\Pi_P=(JP)/(\gamma^2\omega)$, we can introduce
\begin{equation}
 \partial\Pi_{i}=\frac{J\partial P_i(t)}{\gamma^2\omega}~.
\end{equation}
Thus, 
       we can write   the phase deviation $\alpha_i(t)$  as a generalized Fourier series
          by writing its time dependence as a Fourier integral, and expanding 
           its spatial dependence  in terms of the Eigenvectors $ {\bf \phi}_n$ of
           the  generalized Laplace operator 
           $\Lambda$, defined by 
        $\Lambda {\bf \phi}_n = \Lambda_n {\bf \phi}_n$, where
            $\Lambda_n$ are its Eigenvalues\cite{motter,Kettemann2016,coletta}. 
            Thereby we obtain\cite{Kettemann2016,torres}
             \begin{equation}
             \alpha_i(t) = \int_{-\infty}^{\infty} d \epsilon \sum_n c_{n \epsilon} \phi_{ni} e^{-i \epsilon t}.
  \end{equation}            
  Expanding the disturbance likewise in a generalized Fourier series we get 
   \begin{equation}
             \delta \Pi_i(t) = \int_{-\infty}^{\infty} d \epsilon \sum_n \eta_{n \epsilon} \phi_{ni} e^{-i \epsilon t}.
  \end{equation}  
         Now,   we can insert both expansions into Eq. M(9) and find, 
            requiring that the equation 
             is fullfilled for 
            each term of the Fourier series,
            \begin{equation}
            \left( - \tau^2 \epsilon^2 - \i 2 \tau \epsilon + \Lambda_n 
            \right) c_{n \epsilon} = \eta_{n \epsilon}.
            \end{equation} 
             For given disturbance, the  Fourier component of 
              the phase deviation $c_{n \epsilon}$  is thus given 
              in response to the one of the disturbance
                $\eta_{n \epsilon}.$  Inserting that 
                 expression for  $c_{n \epsilon}$ back into the Fourier series we
                  get 
                  \begin{equation}
                  \alpha_i(t) = \int_{-\infty}^{\infty} d \epsilon \sum_n 
                  \left( - \tau^2 \epsilon^2 - \i 2 \tau \epsilon + \Lambda_n  \right)^{-1}
                  \eta_{n \epsilon} \phi_{n i} e^{-i \epsilon t}.
              \end{equation} 
            The integral over the angular frequency $\epsilon$ can be performed 
             by means of the residuum theorem, noting that there are two poles in the lower
              complex plane, $\epsilon_{n \pm} = - \i  (1 \pm \sqrt{1-\Lambda_n}) 1/\tau.$
              We note that the coefficients   $\eta_{n \epsilon}$ are complex and depend on 
              $\epsilon$. If the disturbance sets in at time $t_0$,  $\eta_{n \epsilon}$ has 
              a phase factor $\exp (\i \epsilon t_0).$
              Thus, for $t> t_0$, the integrand is convergent in the lower complex plane,
               so that we can close the integration contour there, and 
              obtain 
                   \begin{eqnarray} \label{linearresponse}
                  \begin{split}
                  \alpha_i(t>t_0) &= -  \frac{\pi}{\tau} \sum_n  \phi_{n i} \frac{1}{\sqrt{1-\Lambda_n}}
              \\ 
              &\quad \times   \left( \eta_{n \epsilon_{n +}} e^{-i \epsilon_{n+} t} - \eta_{n \epsilon_{n-}} 
                 e^{-i \epsilon_{n -} t} \right).
              \end{split}
              \end{eqnarray} 
              For $t< t_0$ the integrand is convergent in the upper complex plane, 
               where there are no poles, so that the residuum is vanishing and we find 
               $  \alpha_i(t<t_0) = 0$. Eq. (\ref{linearresponse}) is valid in linear order  for any
               perturbation $\delta \Pi_i (t)$ and any electricity grid, inserting the 
               Eigenvalues $\Lambda_n$ and Eigenvector components $\phi_{n i}$ of 
                the respective Laplacian. 
               
         For a local perturbation at a site $j$ lasting only a short time interval
          $\Delta t \ll \tau$ around time $t_0$, we can choose the perturbation as
           $\delta \Pi_i (t) = \delta \Pi \delta_{i j} \tau \delta (t-t_0).$         
         Fourier transformation gives $\eta_{n \epsilon} = \frac{1}{2 \pi} \delta \Pi \tau \phi_{n j}^* e^{\i \epsilon t_0}.$ Insertion into Eq. (\ref{linearresponse}) gives then 
         
         \begin{eqnarray} \label{linearresponsedelta}
                 \begin{split}
                   \alpha_i(t>t_0) &= -  \frac{\delta \Pi}{2} \sum_n  \phi_{n i} \phi_{n j}^* \frac{1}{\sqrt{1-\Lambda_n}} \\ 
                   &\quad \times 
                 \left(  e^{-i \epsilon_{n +} (t-t_0)} - e^{-i \epsilon_{n -} (t-t_0)} \right).
              \end{split}
              \end{eqnarray} 
              Thus, it remains to find the Eigenvalues $\Lambda_n$ and Eigenvector components $\phi_{n i}$.
 
 {\it Square Grid.} For  a square grid 
   with transmission line  length $a$ and 
 power capacitance $K$ and periodically 
   arranged generator and consumer power $P_i =\pm P$, 
     the Eigenvectors are plain waves with
       $\phi_{q_n i} = c_{q_n}e^{\imath {\bf q_n r}_i},$ where ${\bf q_n}$ is the wave vector
       which takes on a grid of finite size $L$ discrete values, ${\bf q_n} = 
    (n_x, n_y)     \pi/L,$  where $n_x,n_y \in \{- L/(2a), ... , +L/(2a) \}$. 
       The      Eigenvalues of the Laplacian   $\Lambda$ are
        given by $\Lambda_n = \tau^2 \epsilon_{q_n}^2$, where 
     the Eigenfrequency  $\epsilon_q$ of the linear wave equation is\cite{Kettemann2016},
 \begin{eqnarray} \label{alphaharmonic2}
   \epsilon_q  = \sqrt{ \Pi_K}  (1-\sigma^2/\sigma_c^2)^{1/4} \sqrt{ 4 -f_q} \Gamma_0,
 \end{eqnarray}
 where  
  $\sigma/\sigma_c = P/(4 K)$ and $f_q =2 (\cos q_x a + \cos q_y a)$. 
Insertion in Eq.   (\ref{linearresponsedelta}) thus yields the phase pertubation 
 in response to a  change of the power at time $t_0$ at site $j$, $\delta \Pi_i (t)$ as
 		\begin{eqnarray} \label{linearresponsesl}
		 \begin{split}
                  \alpha_i(t>t_0) &=  \delta \Pi   \frac{1}{N} e^{-(t-t_0)/\tau} \sum_{n_x,n_y}
                   e^{\i {\bf q_n (r_i - r_j)}} 
                   \\ 
                   &\quad \times 
                   \frac{\sinh \left( \frac{t-t_0}{\tau} \sqrt{1-\Lambda_n}
                   \right)}{\sqrt{1-\Lambda_n}}
                   .
                   \end{split}
              \end{eqnarray} 
              Thus, inserting all Eigenvalues $\Lambda_n = \tau^2 \epsilon_{q_n}^2$ 
               with Eq. (\ref{alphaharmonic2})  we get the transient behavior of
                the phase deviation for all times $t> t_0$. 
                    For large momenta $q$, the ballistic limit, the 
  relaxation is fast with the local rate $\Gamma_0$ and there is a real frequency with 
 linear dispersion,  $\Omega_q|_{q\ll0} \approx -\imath \Gamma_0 + v_0 q$ with velocity $v_0 = \sqrt{\Pi_K \cos \delta_0} a/\tau$.
                    At large time $t \gg \tau$ all such modes with  Eigenvalues $\Lambda_n  > 1$
               have decayed. 
               
                Depending on the system 
                   parameters there  can  exist  slow modes with $\Lambda_n  < 1,$
                    which decay with smaller rate 
                    $\Gamma_{n -} =  \Gamma_0  ( 1 -   (1 - \Lambda_n)^{1/2})$. 
  If this condition is fulfilled,    Eigenmodes with  small wave number $q$ appear  whose Eigenfrequency is purely imaginary, 
   $\Omega_q =- \imath  \Pi_K  \cos \delta_0  a^2 { \bf q}^2,$ which decay slowly without oscillations. 
Then, 
  summing over all modes in the spectral representation of $\alpha_i(t)$,
   we find that  an initially localized perturbation at node $l$ spreads 
 for  times $ t> \tau$ and distances exceeding the mean free path $l = v_0 \tau$,
  $|r_{\rm i} -r_{\rm l}| > l$,
  using the continuum limit
\begin{equation}
 \sum_{\vec{q}}\rightarrow\frac{1}{\left(\frac{2\pi}{L}\right)^2}\int\text{d}\vec{q}~~,
\end{equation}
   according to
\begin{eqnarray}
\begin{split}
 \alpha_i(t) &= \delta\Pi\frac{1}{N}\frac{L^2}{4\pi^2}\int_{-\pi/a}^{+\pi/a}\text{d}k_x\int_{-\pi/a}^{+\pi/a}\text{d}k_y 
	\\ &\quad \times e^{\imath(k_x \hat{e}_{ijx}a+k_y\hat{e}_{ijy}a)}
  e^{-(t-t_0)D(k_x^2+k_y^2)},
  \end{split}
\end{eqnarray}
yielding
 \begin{eqnarray} \label{diffusion}
   \alpha_i (t) =\frac{ \alpha_0 a^2}{4 \pi D_0 t} \exp \left( - \frac{({\bf r_{\rm i} -r_{\rm l}})^2}{4 D_0 t} \right).
 \end{eqnarray} 
Thus, the initially localized perturbation spreads diffusively with diffusion constant 
\begin{equation} \label{diffusionconstant}
D =  v_0^2 \tau/2 =  \Pi_K
   (1- \sigma^2/\sigma_c^2)^{1/2} a^2/(2 \tau).
   \end{equation} Diffusion causes very  slow power law relaxation
     of the disturbance at the initial site, and an initial increase, followed by a slow power law decay at other sites.
          The resulting power law relaxation of the change in   transmitted power between 
            nodes $k$ and $l$ is then obtained to be \cite{Kettemann2016}
           \begin{equation} \label{changeFdynamic}
           \delta F_{kl}(t) = \pm  \delta P A_{kl} \frac{\pi^2 a^2  }{\omega_0 D t^2} \exp \left(-\frac{({\bf r_{\rm i} -r_{\rm l}})^2}{4 D t}\right).
           \end{equation}

\end{appendix}

\end{document}